\documentclass[aps,prb,twocolumn,showpacs]{revtex4-1}

\usepackage{graphicx}
\usepackage{dcolumn}
\usepackage{bm}
\usepackage{amsmath}
\usepackage{amssymb}
\usepackage{hyperref}
\usepackage{epstopdf}
\usepackage{caption}

\begin{document}
\title{Out of equilibrium electrons and the Hall conductance of a Floquet topological insulator}
\author{Hossein Dehghani$^{1}$}
\author{Takashi Oka$^{2}$}
\author{Aditi Mitra$^{1}$}
\affiliation{$^1$ Department of Physics, New York University, 4 Washington Place, New York, NY 10003, USA
\\
$^2$ Department of Applied Physics, University of Tokyo, Hongo 7-3-1, Bunkyo, Tokyo 113-8656, Japan
}
\date{\today}

\begin{abstract}
Graphene irradiated by a circularly polarized laser has been predicted to be a Floquet topological insulator showing a laser-induced
quantum Hall effect. A circularly polarized laser also drives the system out of equilibrium resulting in non-thermal
electron distribution functions that strongly affect transport properties. Results are presented
for the Hall conductance for two different cases. One is for a
closed system such as a cold-atomic gas where transverse drift due to non-zero
Berry curvature can be measured in time of flight measurements. For this case the effect of a
circularly polarized laser that has been switched on suddenly is studied. The second
is for an open system coupled to an external reservoir of phonons. While for the
former, the Hall conductance is far from the quantized limit, for the latter, coupling
to a sufficiently low temperature reservoir of phonons is found to produce effective cooling, and thus an approach to the
quantum limit, provided the frequency of the laser is large as compared to the band-width. For laser frequencies comparable to
the band-width, strong deviations from the quantum limit of conductance is found even for a very low temperature
reservoir, with the precise value of the Hall conductance determined by a competition between
reservoir induced cooling and the excitation of photo-carriers by the laser.
For the closed system, the electron distribution function is determined by the overlap between the initial wavefunction
and the Floquet states which can result in a Hall conductance which is opposite in sign to that of the open system.
\end{abstract}

\maketitle

\section{Introduction}

A corner stone in condensed matter has been the discovery of the quantum Hall effect~\cite{Klitzing80,qherev} where
electrons confined to two-dimensions ($2D$) and subjected to an external magnetic field exhibit transport properties
that are remarkable in their insensitivity to material parameters. In particular for the case of the integer quantum
Hall effect, the Hall conductance ($\sigma_{xy}$) is quantized in integer multiples of the universal conductance $e^2/h$
(i.e., $\sigma_{xy}=Ce^2/h$) with the integer $C$ being
a geometric or topological property of the band-structure, known as
the Chern number.~\cite{TKNN,Bellissard94,Avron94} Not surprisingly, the discovery of this effect has lead to tremendous interest in exploring similar
topologically protected transport in other systems.
An important contribution in this direction was the theoretical
proposal of the quantum Hall effect in the absence of a magnetic field, but in the presence of a
staggered magnetic flux
which still breaks time-reversal
symmetry.~\cite{Haldane88} Soon
after, topologically protected transport in 2D and 3D in time-reversal preserving systems was
discovered.~\cite{Hasan10,Zhang11,Kane05,Bernevig06} There is also now a growing interest in generalizing
these concepts to strongly interacting systems.~\cite{Senthil14}

Another intriguing class of systems are those that show topological behavior only dynamically, an example
of this are the
Floquet topological insulators (TIs) where a time-periodic perturbation
modifies the electron hopping matrix elements in such a way as to mimic
a magnetic flux.~\cite{Oka09,Inoue10,Kitagawa10,Lindner11}
Since time-dependent Hamiltonians do not conserve energy, the concept of energy-levels
do not exist. For the particular case of time-periodic Hamiltonians, a quasi-energy spectrum may still be constructed
from the eigenvalues of the time-evolution operator over one period.~\cite{Shirley65,Sambe73}
In this language, Floquet TIs have bulk quasi-energy bands with non-zero
Berry-curvature and Chern number, and support edge-states in confined geometries.
~\cite{Oka09,Inoue10,Kitagawa10,Lindner11,Kitagawa11,Lindner13,Gomez13,Podolsky13,PerezPisk14,Erhai14,Kundu14,Quelle14}

However there are many open questions in the study of Floquet TIs that are unique to the fact that these systems
are out of equilibrium. Firstly,  much of the discussion
in the literature assumes that these quasi-energy levels play the same role as the true energy levels
of a static Hamiltonian, which leads to theoretical predictions of quantum Hall-like quantized transport,~\cite{Oka09,Kitagawa11}
with strong experimental signatures of robust chiral edge transport in optical waveguides.~\cite{Segev13}
However in a nonequilibrium system the electron distribution function, which enters in all
measurable quantities, is not known apriori and depends sensitively on relaxation mechanisms,~\cite{Fertig11,Kundu13,Torres14,Dehghani14,Shirai14,Chamon14}
and at least
on shorter time-scales, on how the external periodic drive has been switched on.~\cite{Lazarides14,Sentef14,Dalibard14,Dehghani14,Rigol14,Burkov14}
Moreover,
unlike static Hamiltonians, there may not even be a one to one correspondence between the Chern number of
the bulk quasi-bands and the number of edge-states in the quasi-spectrum,~\cite{Erhai14} and hence some new topological invariants may be
necessary for time-periodic systems.~\cite{Rudner13,Carpentier14}
Often dissipative coupling to suitably chosen reservoirs can strongly modify
topological properties~\cite{Diehl11,Budich14} thus requiring new measures for
topological order in open and dissipative systems.~\cite{Uhlmann86,Delgado13,Delgado14a}

Understanding these issues is particularly important due to several experimental realizations
of Floquet systems such as in optical waveguides,~\cite{Segev13} cold atoms in
periodically modulated optical lattices,~\cite{Esslinger14} 2D Dirac fermions
on the surface of a 3D TI irradiated by a circularly polarized laser,~\cite{Gedik13,Onishi14}
and chiral transport in graphene irradiated by THz radiation.~\cite{Karch10,Karch11}

In this paper we study graphene irradiated by a circularly polarized laser, taking into account the
full time-evolution of the system, and also accounting for coupling to an external reservoir of phonons.
A similar study was carried out for 2D Dirac fermions~\cite{Dehghani14} where it was shown that
in the absence of coupling to an external reservoir, {\sl i.e.}, when the system was an ideal closed
quantum system, the electron distribution function retained memory of the state before the laser was switched on,
and also depended on the laser switch-on protocol. It was also shown that coupling to phonons makes
the system lose memory of these initial conditions, yet the electron
distribution function was still far out of equilibrium even when the phonons were an ideal reservoir. The effect of
the electron distribution function on the photoemission spectra was discussed.

In this paper our goal is
to study the effect of the electron distribution function on the dc Hall conductance
both for an ideal closed quantum system, and for an
open system. A computation of the Hall conductance requires going beyond the continuum model of Dirac fermions
because the Berry curvature for a Floquet system becomes mathematically ill-defined in the continuum,
in the vicinity of $k$-points where laser induced inter-band transitions are allowed. On a lattice on the other hand,
even in the presence of resonances, the Berry-curvature remains well defined.
Thus in this paper we generalize the treatment of Ref.~\onlinecite{Dehghani14}
to graphene with the aim of exploring the
dc Hall conductance.

Usually Hall conductance is measured in solid-state systems using four terminals or leads, two for driving the current,
and two transverse leads across which the voltage is measured.~\cite{Dattabook}
However in cold-atomic gases one may study the Hall conductance even without leads, by applying a small potential gradient, and
studying the transverse drift of the particles in time of flight measurements.~\cite{Esslinger14} Thus, our results for
the closed system is applicable for such a set-up. Our results for the open system is more relevant to a solid-state
device where the electron-phonon scattering is strong.

We now discuss some subtleties related to transport in two dimensions. In general the conductance and
conductivity are related as conductance =conductivity$\times L^{D-2}$. Thus for $D=2$, both the conductance and conductivity
become independent of the sample size, and a four terminal measurement of the conductance also measures
the conductivity, the latter being typically evaluated within the linear-response Kubo formalism.
At the same time, conductance of mesoscopic systems can also be computed within a Landauer formalism provided there
is no inelastic scattering in the system.~\cite{Dattabook} For larger systems, where electron-electron or
electron-phonon scattering becomes important, the Landauer formalism can no longer be applied.

The Landauer formalism can be generalized to time-periodic systems,~\cite{Hanggi2005} and this approach has been used to compute the
two terminal~\cite{Fertig11,Kundu14} and four terminal~\cite{Torres14} conductance of graphene sheets irradiated by a laser.
This formalism again assumes that there is no inelastic
scattering, and that energy is conserved upto an integer times the laser frequency, with the electron occupation probabilities
primarily determined by the overlap of the Floquet states with the states in the leads.
Our treatment in this paper, employing
the Kubo formalism is in the opposite limit where the sample size is large so that inelastic electron-phonon scattering
is important. Thus our results are in a regime complementary to that addressed in Ref.~\onlinecite{Torres14}. In this limit
of large system size, the mean chemical potential of the leads maintains the average filling (in our case we are always at half filling),
while the voltage difference that maintains current flow is modeled as a small electric field maintained across the sample, and which
is treated within the linear-response Kubo formalism.

The outline of the paper is as follows, in Section~\ref{model} the model is introduced,
a Kubo formula for the dc Hall conductance
is derived, and the ``ideal'' quantum limit explained. In Section~\ref{ideal} the dc Hall
conductance is presented for the closed system and compared with the ``ideal'' case.
In Section~\ref{phonons} we generalize to the open system where the electrons are coupled to a phonon reservoir. The rate or kinetic equation accounting for inelastic electron-phonon scattering in the presence of a periodic drive is derived. The results for the Hall conductance at steady-state with different reservoir temperatures are obtained and compared with results for the closed
system and with the "ideal" case. Finally in section~\ref{conclu} we present our conclusions.

\section{Model} \label{model}
We study graphene irradiated by a circularly polarized laser, and also coupled to a bath of phonons. The Hamiltonian is,
\begin{eqnarray}
H = H_{\rm el} + H_{\rm ph} + H_c
\end{eqnarray}
where (setting $\hbar=1$) $H_{\rm el}$ is the electronic part,
\begin{eqnarray}
&&H_{\rm el}=-t_h\sum_k\begin{pmatrix}c_{kA}^{\dagger}&c_{kB}^{\dagger}\end{pmatrix}
\begin{pmatrix}0&h_{k}^{AB}(t)\\\left[h_{k}^{AB}(t)\right]^*&0\end{pmatrix}
\begin{pmatrix}c_{kA}\\c_{kB}\end{pmatrix},\nonumber\\
&&h^{AB}_k(t) = \sum_{i=1,2,3}e^{ia(\vec{k} + \vec{A}(t))\cdot\vec{\delta}_i}
\end{eqnarray}
$\vec{\delta}_i$ are the nearest-neighbor unit-vectors on the graphene lattice,
$\vec{\delta}_1= \left(\frac{1}{2},\frac{\sqrt{3}}{2}\right);\vec{\delta}_2= \left(\frac{1}{2},-\frac{\sqrt{3}}{2}\right);
\vec{\delta}_3= \left(-1,0\right)$. The circularly polarized laser enters through minimal substitution
$\vec{k}\rightarrow \vec{k} + \vec{A}(t)$, where
\begin{eqnarray}
A_x(t) = \theta(t)A_0 \cos\Omega t; A_y(t) = - \theta(t)A_0\sin\Omega t
\end{eqnarray}
We assume that the laser has been suddenly switched on at time $t=0$. This assumption holds equally well
for lasers switched on over a time which is short as compared to the period $2\pi/\Omega$ of the laser.

The translation vectors $\vec{a}_1,\vec{a}_2$ for graphene are
$\vec{a}_1 = \frac{a}{2}\left(3,\sqrt{3}\right); \vec{a}_2 = \frac{a}{2}\left(3,-\sqrt{3}\right)$, while
the reciprocal lattice vectors ($\vec{b}_i\cdot\vec{a}_j=2\pi \delta_{ij}$) are
$\vec{b}_1 = \frac{2\pi}{3a}\left(1,\sqrt{3}\right); \vec{b}_2 = \frac{2\pi}{3a}\left(1,-\sqrt{3}\right)$.
As written above, $h^{AB}_k$ is not invariant under translations by integer multiples of a reciprocal lattice vector, $\vec{k}\rightarrow \vec{k} + n_i\vec{b}_i$. In order to recover this symmetry it is convenient to
make the transformation $c_{kB}\rightarrow c_{kB}e^{ia\vec{k}\cdot\vec{\delta}_3}$.~\cite{Bernevigbook}
Then since, $a(\vec{\delta}_1-\vec{\delta}_3)=\vec{a}_1$,
$a(\vec{\delta}_2-\vec{\delta}_3)=\vec{a}_2$, after this transformation, $h^{AB}_k$ becomes
\begin{eqnarray}
&&h_k^{AB}(t) = e^{i a \vec{A}(t)\cdot \vec{\delta}_3} +\sum_{i=1,2}e^{i\vec{k}\cdot\vec{a}_i + ia\vec{A}(t)\cdot\vec{\delta}_i}
\end{eqnarray}

Dissipation affects the electron distribution
and thus the topological signatures such as the Hall conductance.
Here we consider dissipation due to coupling to 2D phonons
\begin{eqnarray}
H_{\rm ph}=\sum_{q,i=x,y}\left[\omega_{qi}b_{qi}^{\dagger}b_{qi}\right]
\end{eqnarray}
where the electron-phonon coupling is
\begin{eqnarray}
&&H_c = \sum_{{k}q\sigma,\sigma'=A,B}
c_{k\sigma}^{\dagger}\vec{A}_{\rm ph}(q)\cdot\vec{\sigma}_{\sigma \sigma'}c_{{k}\sigma'}\\
&&\vec{A}_{\rm ph}(q) = \left[\lambda_{x,q}\left(b_{x,q}^{\dagger}+b_{x,-q}\right), \lambda_{y,q}\left(b_{y,q}^{\dagger}+b_{y,-q}\right)
\right]
\end{eqnarray}
As is standard practice, we have denoted the sub-lattice labels $A,B$ in terms of a pseudo-spin label $\sigma$, a notation
that will be adopted throughout the paper.
Above we have made the assumption that phonon induced
scattering between electrons with different quasi-momenta do not occur.
Thus electronic states at different quasi-momenta $k$ are independently coupled to
the reservoir.

Moreover we will later assume that the phonons have a broad bandwidth
so that
inelastic scattering channels between electrons at all relevant quasi-energies and phonons are possible.
While electron quasi-energies form an infinite ladder which may be regarded as
photon absorption and emission side-bands, the matrix elements between different quasi-energy bands
and phonons are suppressed very rapidly as the number of photon absorption and emission
processes increase~\cite{Dehghani14}. Thus
for the laser amplitude and frequencies we will be working with, to have the most effective inelastic relaxation,
it will be sufficient to consider
a maximum phonon frequency $\omega_q^{\rm max}\simeq 6\Omega$.
A circularly polarized laser also opens up a gap $\Delta$ at the Dirac points which in the high frequency
limit of $A_0at_h/\Omega \ll 1$
is $\Delta \simeq 2A_0^2a^2t_h^2/\Omega$.~\cite{Oka09,Kitagawa11} Thus we will assume that the lowest phonon frequency
available is $\omega_q^{\rm min} \simeq \Delta$ to allow for efficient relaxation near the Dirac points.

\subsection{Kubo formula for the Hall conductance}\label{kubo}

The Kubo formula for the Hall conductance is a linear response to a weak
probe $\vec{A}_{\rm pr}$ that is applied over and above
the circularly polarized laser $\vec{A}$. While the Kubo formalism has been employed before for Floquet
systems~\cite{Torres05,Oka09,Oka10}, yet we outline the derivation in order to highlight the
main assumptions, and also in order to generalize the derivation to open systems
like the one studied in this paper.

The electronic part of the Hamiltonian in the presence of an external laser $\vec{A}$ and a
probe field $\vec{A}_{\rm pr}$ is,
\begin{eqnarray}
H_{\rm el}'=\sum_{ij\sigma\sigma'}c_{i\sigma}^{\dagger}h_{ij}^{\sigma\sigma'}c_{j\sigma'}e^{-i\int_{j}^i\left(\vec{A}(t)+\vec{A}_{\rm pr}(t)\right)\cdot\vec{dl}}
\end{eqnarray}
Since $c_{j\sigma} =\frac{1}{\sqrt{N}}\sum_ke^{i \vec{k} \cdot\vec{j}}c_{k\sigma}$, we see that the vector potential corresponds to
replacing $\vec{k} \rightarrow \vec{k} + \vec{A} + \vec{A}_{\rm pr}$.
Taylor expanding with respect to the weak probe,
\begin{eqnarray}
&&H_{\rm el}'\simeq  \sum_{jr\sigma\sigma'}c_{j+r,\sigma}^{\dagger}h_{j+r,j}^{\sigma\sigma'}(t)c_{j\sigma'}\left(1-i\int_{j}^{j+r}\vec{A}_{\rm pr}(t)\cdot\vec{dl}\right)
\nonumber\\
&&= H_{\rm el}+\sum_{q}\vec{j}_q\cdot\vec{A}_{-q}^{\rm pr}
\end{eqnarray}
where $\vec{A}_{\rm pr}(\vec{j}) = (1/\sqrt{N})\sum_qe^{i \vec{q}\cdot \vec{j}}\vec{A}_{q}^{\rm pr}$ and
\begin{eqnarray}
\vec{j}_q=\frac{1}{\sqrt{N}}\sum_{k,\sigma\sigma'}c_{k+q/2,\sigma}^{\dagger}c_{k-q/2,\sigma'}\frac{\partial h_k^{\sigma \sigma'}(t)}{\partial \vec{k}}
\end{eqnarray}
The current-current correlation function which quantifies how an electric field applied in the direction $\hat{i}$,
affects the current flowing in the direction $\hat{j}$ is given by
\begin{eqnarray}
&&R_{ij}(q,t,t')=-i\theta(t,t')\biggl\langle \Psi(t_0)\biggl |\biggl[j_{qI}^i(t),j_{-qI}^j(t')\biggr]\biggr |\Psi(t_0)\biggr\rangle\nonumber\\
\end{eqnarray}
where $|\Psi(t_0)\rangle$ is the wavefunction at a certain reference time $t_0$, while the current operators are in the
interaction representation
\begin{eqnarray}
&&\vec{j}_{kI}(t) = U_k(t_0,t)\vec{j}_k U_k(t,t_0)
\end{eqnarray}
where $U_k(t,t')$ is the time-evolution operator due to the electronic part of the
Hamiltonian ($H_{\rm el}$), and is given by
\begin{eqnarray}
U_k(t,t_0)=\sum_{\alpha=u,d}e^{-i\epsilon_{k\alpha}(t-t_0)}|\phi_{k\alpha}(t)\rangle\langle\phi_{k\alpha}(t_0)|\label{Udef}
\end{eqnarray}
Above $\epsilon_{k\alpha=u,d}$ are the quasi-energies while
\begin{eqnarray}
|\phi_{k\alpha}(t)\rangle=\begin{pmatrix}\phi_{k\alpha}^{\uparrow}\\ \phi_{k\alpha}^{\downarrow}\end{pmatrix}
\end{eqnarray}
are the quasi-modes that are periodic in time. Thus,
$U_{k\sigma\sigma'}(t,t_0)= \sum_{\alpha}e^{-i\epsilon_{k\alpha}(t-t_0)}\phi_{k\alpha}^{\sigma}(t)\phi_{k\alpha}^{\sigma'*}(t_0)$, and
in the interaction representation $c_{k\sigma}(t) = U_{k\sigma\sigma'}(t,t_0)c_{k\sigma'}(t_0)
,c_{k\sigma}^{\dagger}(t) = c_{k\sigma'}^{\dagger}(t_0) U_{k\sigma'\sigma}(t_0,t)$.
The quasi-energies $\epsilon_{k\alpha}$ represent an infinite ladder of
states where $\epsilon_{k\alpha}$ and $\epsilon_{k\alpha}+m\Omega$, for any integer $m$,
represent the same physical state corresponding to the
Floquet quasi-modes $|\phi_{k\alpha}(t)\rangle$ and
$e^{im\Omega t}|\phi_{k\alpha}(t)\rangle$ respectively. Confusion due to this over-counting can be easily avoided by noting that
in all physical quantities and matrix elements, it is always the combination $e^{-i\epsilon_{k\alpha}t}|\phi_{k\alpha}(t)\rangle
= |\psi_{k\alpha}\rangle$ that appears, where $|\psi_{k\alpha}\rangle$
are the solutions to the time-dependent Schrodinger equation. There are only two distinct solutions for $|\psi_{k\alpha}\rangle$
which we label as $\alpha=u,d$, while we adopt the convention that the corresponding quasi-energies lie within
a Floquet Brillouin zone (BZ) $-\Omega/2<\epsilon_{k\alpha}<\Omega/2$.

Expanding the fermionic operators in the quasi-mode basis at $t_0$,
\begin{eqnarray}
c_{kb'}(t_0) = \sum_{\alpha'}\phi_{k\alpha'}^{b'}(t_0)\gamma_{k\alpha'}
\end{eqnarray}
where $\gamma^{\dagger}_{k\alpha},\gamma_{k\alpha}$ are the creation and annihilation operators for
the quasi-modes at time $t_0$,
the response function at $q=0$ is found to be
\begin{eqnarray}
&&R_{ij}(q=0,t,t')= -i\theta(t-t')\frac{1}{N}\sum_{k,\alpha\beta\gamma\delta}\nonumber\\
&&e^{-i(\epsilon_{k\alpha}-\epsilon_{k\beta})(t-t_0)}
e^{-i(\epsilon_{k\gamma}-\epsilon_{k\delta})(t'-t_0)}\nonumber\\
&&\langle\phi_{k\beta}(t)|\biggl[\frac{\partial h_k(t)}{\partial {k}_i}\biggr]|\phi_{k\alpha}(t)\rangle
\langle\phi_{k\delta}(t')|\biggl[\frac{\partial h_k(t')}{\partial {k}_j}\biggr]|\phi_{k\gamma}(t')\rangle\nonumber\\
&&
\biggl\langle\Psi(t_0)\biggl|\biggl[\gamma^{\dagger}_{k\beta}\gamma_{k\alpha},\gamma^{\dagger}_{k\delta}\gamma_{k\gamma}\biggr]
\biggr|\Psi(t_0)\biggr\rangle
\end{eqnarray}
Since the Floquet quasi-modes at any given time form a complete basis that obey
\begin{eqnarray}
\left[h_k-i\partial_t\right]|\phi_{k\alpha}\rangle = \epsilon_{k\alpha}|\phi_{k\alpha}\rangle
\end{eqnarray}
the following relation holds,
\begin{eqnarray}
&&\langle \phi_{k\beta}|\nabla h_k|\phi_{k\alpha}\rangle =i \partial_t\left[\langle \phi_{k\beta}|\nabla\phi_{k\alpha}\rangle\right]
+  \delta_{\alpha\beta}\nabla\epsilon_{k\alpha} \nonumber\\
&&+ \left(\epsilon_{k\alpha}-\epsilon_{k\beta}\right)\langle \phi_{k\beta}|\nabla\phi_{k\alpha}\rangle
\end{eqnarray}
$R_{ij}(t,t')$ depends not only on the time difference $t-t'$ but also on the mean time $(t+t')/2$. In what follows
we will make some approximations that are equivalent to averaging over the mean time.

The first approximation we make is to retain only diagonal components of the average below, since the
off-diagonal terms will be accompanied by oscillations of the kind $e^{-i(\epsilon_{ku}-\epsilon_{kd})(t+t')/2}$,
\begin{eqnarray}
\biggl\langle\biggl[\gamma^{\dagger}_{k\beta}\gamma_{k\alpha},\gamma^{\dagger}_{k\delta}\gamma_{k\gamma}\biggr]\biggr\rangle
&&= \delta_{\alpha,\delta}\delta_{\beta,\gamma}\biggl[\langle \gamma_{k\beta}^{\dagger}\gamma_{k\beta}\rangle\nonumber\\
&& -\langle
\gamma_{k\alpha}^{\dagger}\gamma_{k\alpha}
\rangle\biggr]
\end{eqnarray}
Thus we obtain,
\begin{eqnarray}
&&R_{ij}(q=0,t,t')= -i\theta(t-t')\sum_{k,\alpha,\beta}e^{-i(\epsilon_{k\alpha}-\epsilon_{k\beta})(t-t')}\nonumber\\
&&\biggl[\left(\epsilon_{k\alpha}-\epsilon_{k\beta}\right)\langle\phi_{k\beta}(t)|
\frac{\partial }{\partial {k}_i}\phi_{k\alpha}(t)\rangle
+ i\partial_t\langle\phi_{k\beta}(t)|\frac{\partial }{\partial {k}_i}\phi_{k\alpha}(t)\rangle\biggr]\nonumber\\
&&\biggl[-\left(\epsilon_{k\alpha}-\epsilon_{k\beta}\right)\langle\phi_{k\alpha}(t')|
\frac{\partial }{\partial {k}_j}\phi_{k\beta}(t')\rangle\nonumber\\
&&+ i\partial_{t'}\langle\phi_{k\alpha}(t')|\frac{\partial }{\partial {k}_j}\phi_{k\beta}(t')
\rangle\biggr]\biggl[\langle \gamma_{k\beta}^{\dagger}\gamma_{k\beta}\rangle-\langle
\gamma_{k\alpha}^{\dagger}\gamma_{k\alpha}
\rangle\biggr]
\end{eqnarray}
Let us define
\begin{eqnarray}
\langle\phi_{k\beta}(t)|\frac{\partial }{\partial {k}_i}\phi_{k\alpha}(t)\rangle = \sum_{m}e^{im\Omega t}
C^m_{\beta i \alpha}
\end{eqnarray}
then,
\begin{eqnarray}
&&R_{ij}(q=0,t,t')= -i\theta(t-t')\sum_{k,\alpha,\beta,m,m'}e^{im\Omega t +i m'\Omega t'}\nonumber\\
&&e^{-i(\epsilon_{k\alpha}-\epsilon_{k\beta})(t-t')}C^m_{\beta i \alpha}C^{m'}_{\alpha j \beta}\nonumber\\
&&
\biggl[\epsilon_{k\alpha}-\epsilon_{k\beta}-m\Omega\biggr]\biggl[-\left(\epsilon_{k\alpha}-\epsilon_{k\beta}\right)-m'\Omega
\biggr]\nonumber\\
&&\times \biggl[\langle \gamma_{k\beta}^{\dagger}\gamma_{k\beta}\rangle-\langle
\gamma_{k\alpha}^{\dagger}\gamma_{k\alpha}
\rangle\biggr]
\end{eqnarray}
Now we average over one the mean time $(t+t')/2$ over one cycle of the laser. This is equivalent to keeping
only $m'=-m$ terms so that the results become time-translationally invariant,
\begin{eqnarray}
&&R_{ij}(q=0,t,t')= -i\theta(t-t')(-1)\sum_{k,\alpha,\beta,m}e^{im\Omega t -i m\Omega t'}\nonumber\\
&&\times e^{-i(\epsilon_{k\alpha}-\epsilon_{k\beta})(t-t')}C^m_{\beta i \alpha}
C^{-m}_{\alpha j \beta}\nonumber\\
&&
\biggl[\epsilon_{k\alpha}-\epsilon_{k\beta}-m\Omega\biggr]^2
\biggl[\langle \gamma_{k\beta}^{\dagger}\gamma_{k\beta}\rangle-\langle
\gamma_{k\alpha}^{\dagger}\gamma_{k\alpha}
\rangle\biggr]
\end{eqnarray}
Denoting $\alpha=u,\beta=d$, and setting $m\rightarrow -m$ in one of the terms, we obtain,
\begin{eqnarray}
&&R_{ij}(q=0,t,t')= -i\theta(t-t')(-1)\sum_{k,m}\biggl[\epsilon_{ku}-\epsilon_{kd}-m\Omega\biggr]^2\nonumber\\
&&\times \biggl(e^{-i(\epsilon_{ku}-\epsilon_{kd}-m\Omega)(t-t')}C^m_{d i u}
C^{-m}_{u j d}\nonumber\\
&&-e^{-i(\epsilon_{kd}-\epsilon_{ku}+m\Omega)(t-t')}C^{-m}_{u i d}
C^{m}_{d j u}\biggr)\biggl[\langle \gamma_{kd}^{\dagger}\gamma_{kd}\rangle-\langle
\gamma_{ku}^{\dagger}\gamma_{ku}
\rangle\biggr]\nonumber\\
\end{eqnarray}
Fourier transforming this expression,
\begin{eqnarray}
&&R_{ij}(q=0,\omega)=\sum_{k,m}\biggl[\epsilon_{ku}-\epsilon_{kd}-m\Omega\biggr]^2\nonumber\\
&&\biggl[\frac{C^{-m}_{u i d}
C^{m}_{d j u}}{\omega + i\delta +\epsilon_{ku}-\epsilon_{kd}-m\Omega}- \frac{C^m_{d i u}
C^{-m}_{u j d}}{\omega + i\delta -(\epsilon_{ku}-\epsilon_{kd}-m\Omega)}\biggr]\nonumber\\
&&\biggl[\langle \gamma_{kd}^{\dagger}\gamma_{kd}\rangle-\langle
\gamma_{ku}^{\dagger}\gamma_{ku}
\rangle\biggr]
\end{eqnarray}
For the Hall conductance, we need the combination,
\begin{eqnarray}
&&R_{ij}(q=0,\omega) - R_{ji}(q=0,\omega)=\sum_{k,m}\biggl[\epsilon_{ku}-\epsilon_{kd}-m\Omega\biggr]^2\nonumber\\
&&\times \biggl(C^{-m}_{u i d}
C^{m}_{d j u}-C^{-m}_{u j d}
C^{m}_{d i u}\biggr)
\frac{2\left(\omega+i\delta\right)}{\left(\omega + i\delta\right)^2 -\left(\epsilon_{ku}-\epsilon_{kd}-m\Omega\right)^2}\nonumber\\
&&\times \biggl[\langle \gamma_{kd}^{\dagger}\gamma_{kd}\rangle-\langle
\gamma_{ku}^{\dagger}\gamma_{ku}
\rangle\biggr]
\end{eqnarray}
Thus the dc Hall conductance is
\begin{eqnarray}
&&\sigma_{ij}(\omega = 0)= \frac{R_{ij}-R_{ji}}{2i\omega}\biggr|_{\omega=0}\nonumber\\
&&= i\sum_{k,m}\biggl(C^{-m}_{u i d}
C^{m}_{d j u}-C^{-m}_{u j d}
C^{m}_{d i u}\biggr)\biggl[\langle \gamma_{kd}^{\dagger}\gamma_{kd}\rangle-\langle
\gamma_{ku}^{\dagger}\gamma_{ku}
\rangle\biggr]\nonumber\\
\end{eqnarray}
Denoting the laser period as $T_{\Omega}= 2\pi/\Omega$,
\begin{eqnarray}
&&i\sum_m\biggl(C^{-m}_{u i d}C^{m}_{d j u}-C^{-m}_{u j d}C^{m}_{d i u}\biggr)=
i\frac{1}{T^2_{\Omega}}\int_0^{T_{\Omega}}dt_1\int_0^{T_{\Omega}} dt_2\sum_m\nonumber\\
&&e^{-im\Omega(t_1-t_2)}
\biggl(\langle\phi_{kd}(t_1)|\frac{\partial }{\partial {k}_j}\phi_{ku}(t_1)\rangle
\langle\phi_{ku}(t_2)|\frac{\partial }{\partial {k}_i}\phi_{kd}(t_2)\rangle \nonumber\\
&&- \langle\phi_{kd}(t_1)|\frac{\partial }{\partial {k}_i}\phi_{ku}(t_1)\rangle
\langle\phi_{ku}(t_2)|\frac{\partial }{\partial {k}_j}\phi_{kd}(t_2)\rangle\biggr)
\end{eqnarray}
Using $\sum_me^{im\Omega t}=\delta(t/T_{\Omega})$, we obtain,
\begin{eqnarray}
&&i\sum_m\biggl(C^{-m}_{u i d}C^{m}_{d j u}-C^{-m}_{u j d}C^{m}_{d i u}\biggr)= \nonumber\\
&&i\frac{1}{T_{\Omega}}\int_0^{T_{\Omega}}dt
\biggl(\langle\phi_{kd}(t)|\frac{\partial }{\partial {k}_j}\phi_{ku}(t)\rangle
\langle\phi_{ku}(t)|\frac{\partial }{\partial {k}_i}\phi_{kd}(t)\rangle \nonumber\\
&&- \langle\phi_{kd}(t)|\frac{\partial }{\partial {k}_i}\phi_{ku}(t)\rangle
\langle\phi_{ku}(t)|\frac{\partial }{\partial {k}_j}\phi_{kd}(t)\rangle\biggr)\nonumber\\
&&=i\frac{1}{T_{\Omega}}\int_0^{T_{\Omega}}dt
\biggl(\langle\partial_i\phi_{kd}(t)|\partial_j\phi_{kd}(t)\rangle- \langle\partial_j\phi_{kd}(t)|\partial_i\phi_{kd}(t)\rangle\biggr)\nonumber\\\
&&= \frac{1}{T_{\Omega}}\int_0^{T_{\Omega}} dt F_{kd}(t)
\end{eqnarray}
where above we have used the orthonormality of the Floquet states at any given time.
Thus the dc Hall conductance is,
\begin{eqnarray}
\sigma_{xy}(\omega=0)= \frac{e^2}{2\pi h}\int_{BZ}d^2k \overline{F}_{kd}\left[\rho_{kd}-\rho_{ku}\right]\label{sig1}
\end{eqnarray}
where $\overline{F}_{kd}$ is the time-average of the Berry curvature over one cycle,
\begin{eqnarray}
\overline{F}_{kd}=\frac{1}{T_{\Omega}}\int_0^{T_{\Omega}}dt\,\,2{\rm Im}\biggl[\langle \partial_y\phi_{kd}(t)|
\partial_x\phi_{kd}(t)\rangle\biggr]
\end{eqnarray}
and as expected, the Hall conductance depends on the occupation probabilities
\begin{eqnarray}
\rho_{k\alpha=u,d}= \langle\gamma_{k\alpha}^{\dagger}\gamma_{k\alpha}\rangle
\end{eqnarray}
The ``ideal'' quantum limit corresponds to the case where $|\rho_{kd}-\rho_{ku}|=1$, so that the Hall conductance is
\begin{eqnarray}
\sigma_{xy}^{\rm ideal} = C \frac{e^2}{h}
\end{eqnarray}
with
\begin{eqnarray}
C = \frac{1}{2\pi}\int_{\rm BZ} d^2kF_{kd}
\end{eqnarray}
being the Chern number. It is important to note that while the Berry-curvature is time-dependent, it's integral
over the BZ is time-independent, and a topological invariant. However, once the population $\rho_{kd}-\rho_{ku}$
becomes dependent on momentum, the integral of the Berry-curvature weighted by the population is no longer a topological
invariant, and depends on time. The averaging procedure outlined above corresponds to replacing the time-dependent Berry
curvature by its average over one cycle.

In this paper we will study the time-averaged dc Hall conductance defined in Eq.~(\ref{sig1}) for two cases. One is when
the occupation probabilities $\rho_{kd,u}$ are for the closed system
with a quench switch-on protocol for the laser (section~\ref{ideal}), while the second
is for the open system coupled to a reservoir, where the $\rho_{kd,u}$
will be determined from solving a kinetic equation (section~\ref{phonons}).
In order to compute the Berry curvature $F_{kd}$, we will employ the numerical approach of Ref.~\onlinecite{Fukui05}.
\begin{figure}
\begin{center}
\begin{tabular}{lll}
 \includegraphics[height=8cm,width=8cm,keepaspectratio]{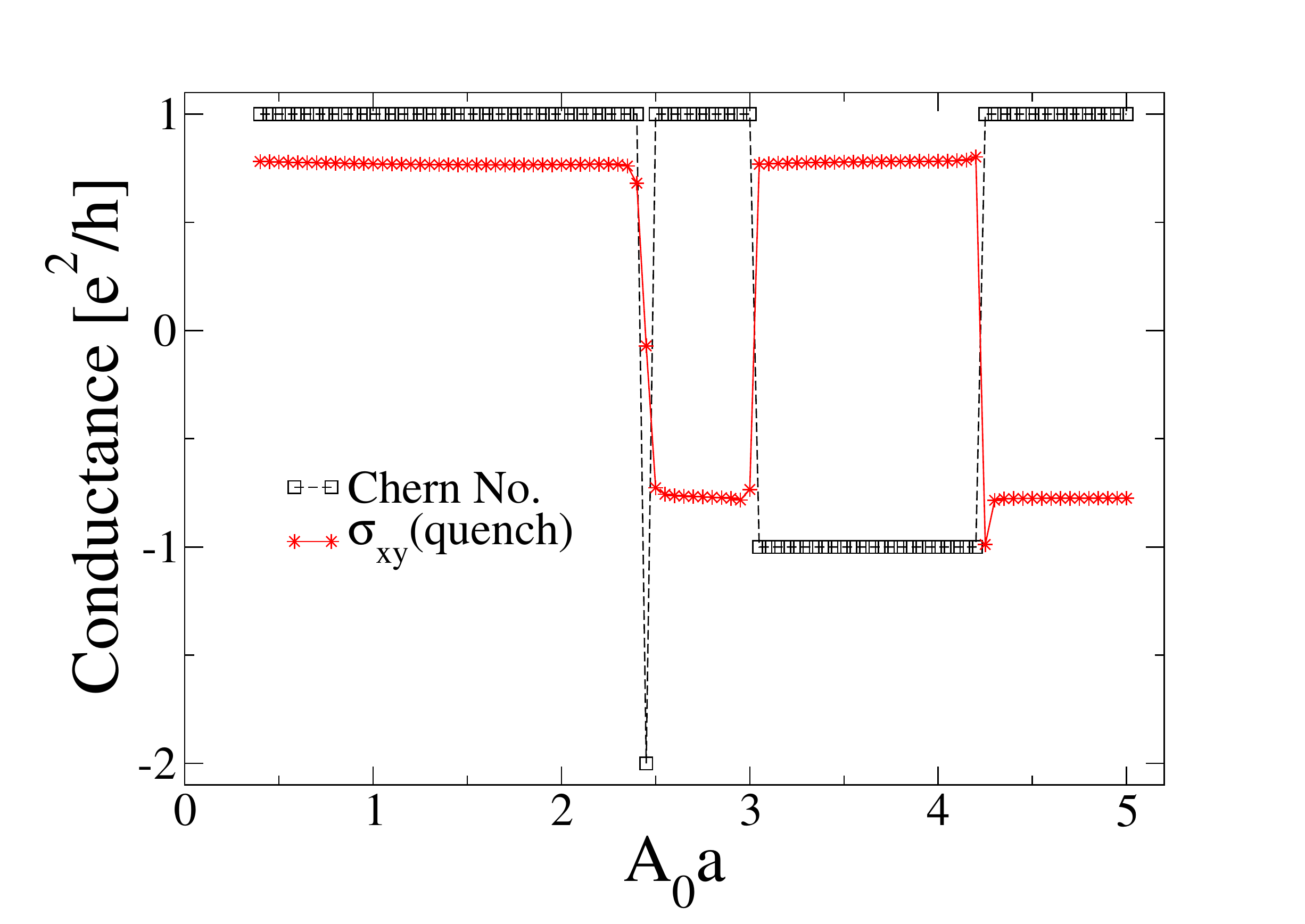}\\
(a)\\
\includegraphics[height=8cm,width=8cm,keepaspectratio]{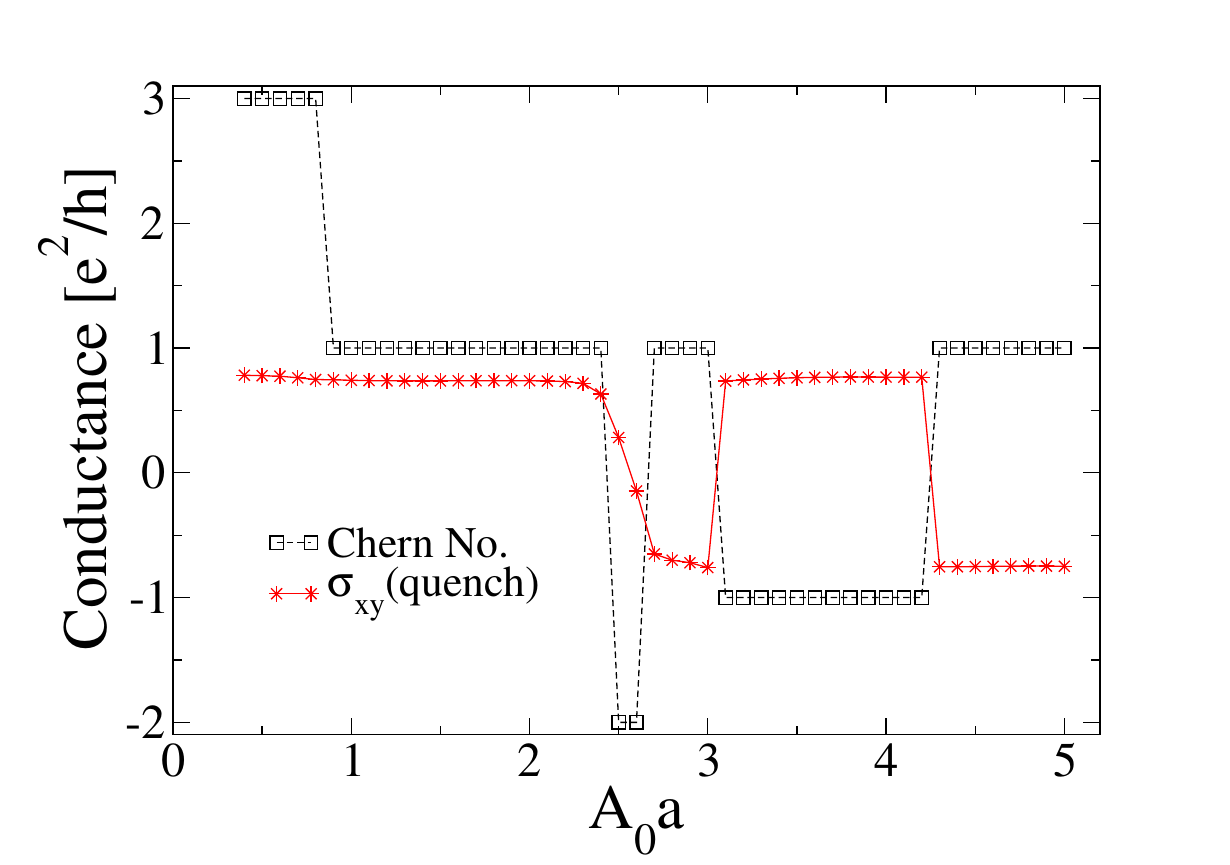}\\
(b)
\end{tabular}
\caption{(Color online) Hall conductance for the ideal case ($\sigma_{xy}= C e^2/h$)
and for a closed system after a quench, for different strengths of the circularly polarized
laser of frequency: (a). $\Omega = 10t_h$
(b). $\Omega=5t_h$.
}
\label{fig1a2}
\end{center}
\end{figure}

\begin{figure}
\begin{center}
\begin{tabular}{lll}
 \includegraphics[height=8cm,width=8cm,keepaspectratio]{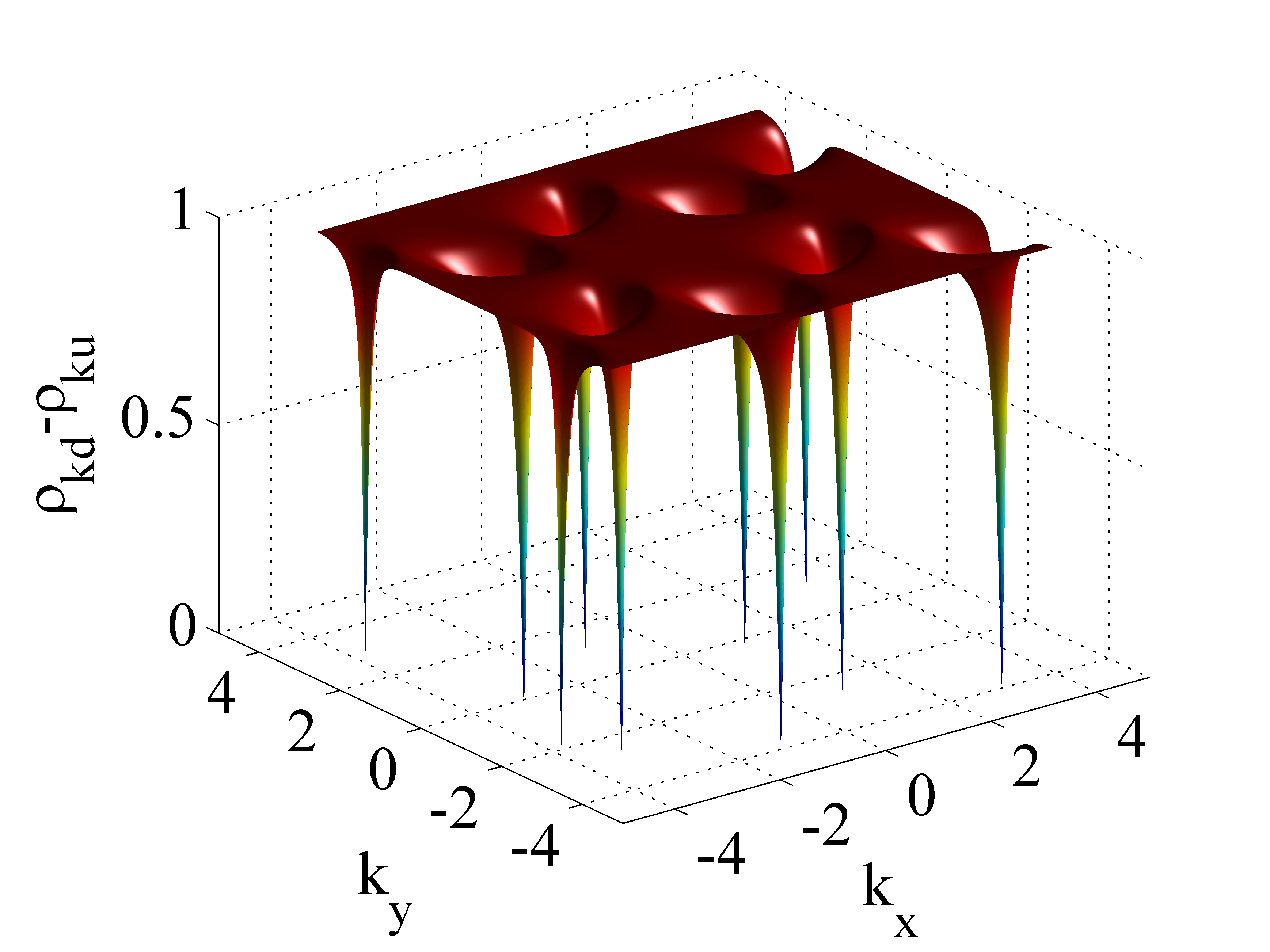}\\
(a)\\
\includegraphics[height=8cm,width=8cm,keepaspectratio]{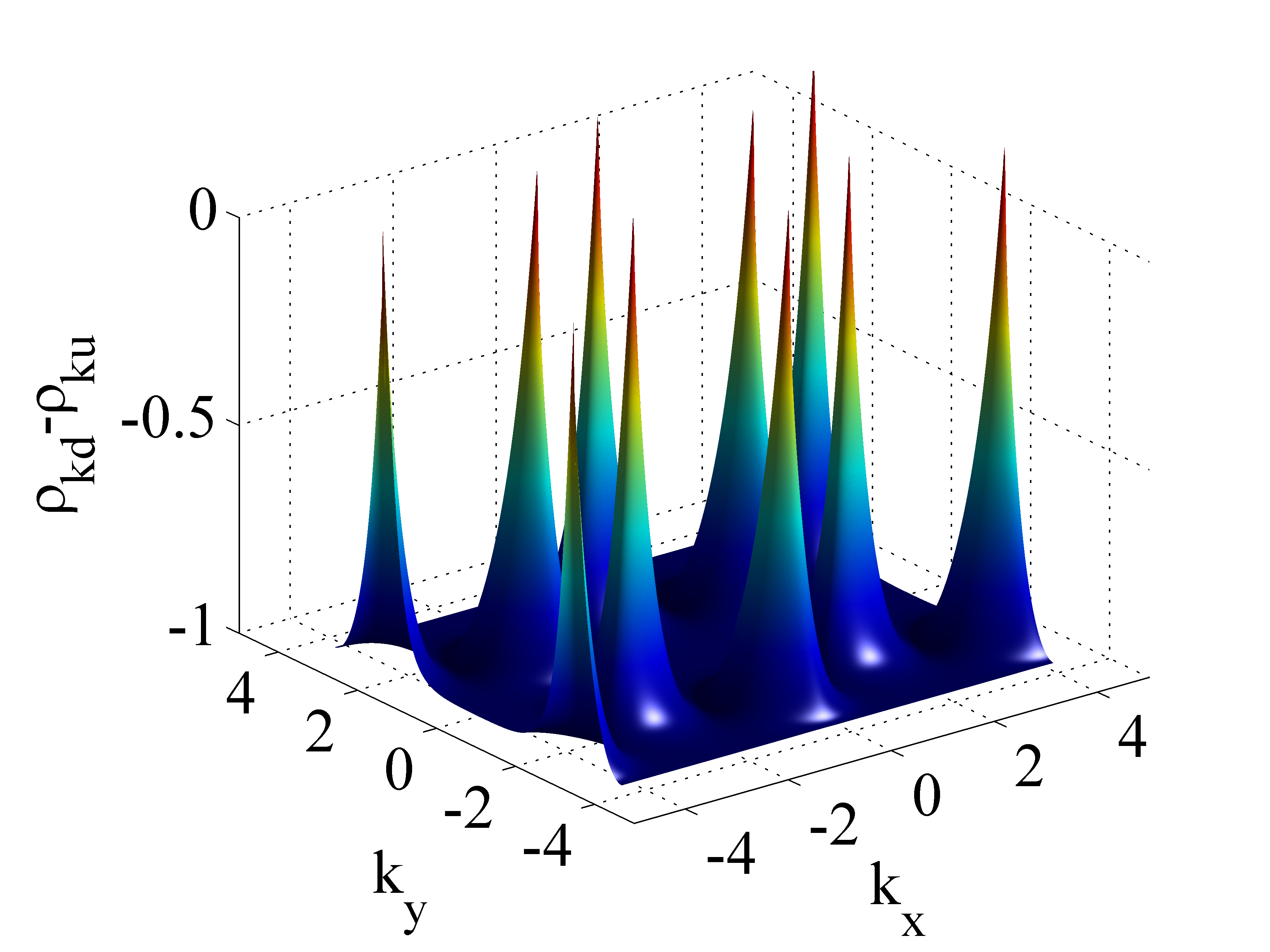}\\
(b)
\end{tabular}
\caption{(Color online) Excitation density $\rho_{kd}-\rho_{ku}$ in the closed system for a quench switch-on protocol
for a laser of frequency $\Omega=10 t_h$ and amplitude:
(a). $A_0a = 1.0$ (b). $A_0a = 5.0$.
}
\label{fig3a4}
\end{center}
\end{figure}

\section{Hall conductance for the closed system for a quench switch-on protocol}\label{ideal}

Suppose that at $t\leq 0$, there is no external irradiation, and the
electrons are in the ground-state of graphene. Thus
the wavefunction right before the switching
on of the laser is
\begin{eqnarray}
&&|\Psi_{\rm in}(t=0^-)\rangle=\prod_{{k}}|\psi_{{\rm in},k}\rangle,\nonumber\\
&&|\psi_{{\rm in},k}\rangle= \frac{1}{\sqrt{2}}
\begin{pmatrix}e^{i\theta_k}\\ 1 \end{pmatrix}
\end{eqnarray}
where
\begin{eqnarray}
&&\tan{\theta_k}= \frac{\sin(\vec{k}\cdot{\vec{a}}_1)
+ \sin{(\vec{k}\cdot{\vec{a}}_2)}}
{1+\cos(\vec{k}\cdot{\vec{a}}_1)
+ \cos{(\vec{k}\cdot{\vec{a}}_2)}}
\end{eqnarray}
The time-evolution after switching
on the laser is
\begin{eqnarray}
|\Psi(t>0)\rangle=\prod_kU_{k}(t,0)|\psi_{\rm in,k}\rangle
\end{eqnarray}
where   $U_{k}(t,t')$ is the time-evolution operator given in Eq.~(\ref{Udef}).

In practice, in order to determine the Floquet states, it is convenient to solve the problem in Fourier space,
\begin{eqnarray}
|\phi_{k\alpha}(t)\rangle =
\sum_m e^{i m \Omega t}|\tilde{\phi}_{m k\alpha }\rangle \label{defft}
\end{eqnarray}
where $|\tilde{\phi}_{mk\alpha}\rangle$ is a 2 component spinor which obeys,
\begin{eqnarray}
&&\sum_m\left[H_{\rm el}^{nm}+ m \Omega\delta_{nm}\right]|\tilde{\phi}_{mk\alpha}\rangle = \epsilon_{k\alpha}|\tilde{\phi}_{nk\alpha}\rangle, \nonumber \\
&&H_{\rm el}^{nm}= \frac{1}{T_{\Omega}}\int_0^{T_\Omega}dt e^{-i(n-m)\Omega t}H_{\rm el}\nonumber\\
&&=\begin{pmatrix}0 & h_{\sigma\sigma'}^{nm}(k) \\ h_{\sigma'\sigma}^{nm}(k) &0 \end{pmatrix}
\label{HF3}
\end{eqnarray}
For graphene in a circularly polarized laser,
\begin{eqnarray}
&&h^{nm}_{\sigma\sigma'}(k)= \!\!-t_hi^{m-n}J_{m-n}\left(A_0a\right)
\sum_{j=1,2,3}\!\!e^{i\vec{k}\cdot\vec{a}_j}e^{-i(m-n)\alpha_j},\nonumber\\
&&h^{nm}_{\sigma'\sigma}(k)\!\!=\!\!
-t_h(-i)^{m-n}J_{m-n}\left(A_0a\right)
\!\!\sum_{j=1,2,3}\!\!\!\!e^{-i\vec{k}\cdot\vec{a}_j}e^{-i(m-n)\alpha_j}\nonumber\\
\end{eqnarray}
where $\alpha_1 = -\alpha_2=\frac{\pi}{3}, \alpha_3=\pi$ and
$\vec{a}_1 = \frac{a}{2}\left(3,\sqrt{3}\right), \vec{a}_2 = \frac{a}{2}\left(3,-\sqrt{3}\right), \vec{a}_3=0$

We are interested in the time-averaged Hall conductance defined in Eq.~(\ref{sig1}). For this we need
the overlap between the initial state before the quench and the Floquet quasi-modes since they control
the occupation probabilities,
\begin{eqnarray}
\rho_{k\alpha=u,d}^{\rm quench} = |\langle \phi_{k \alpha=u,d}(0)|\psi_{\rm in,k}\rangle|^2
\end{eqnarray}

Fig~\ref{fig1a2} shows the Hall conductance for the ideal case where only one Floquet band is occupied
($\rho_{kd}=1,\sigma_{xy}^{\rm ideal}= C e^2/h$), and compared with the Hall conductance
for the quench. Thus each point in the plot corresponds to a situation where initially the
system was in the ground state of graphene, and then at time $t=0$ a laser of strength $A_0a$
and frequency $\Omega$ was switched on suddenly.
Notice that there are a number of topological phase transitions
corresponding to jumps in the Chern number as $A_0at_h/\Omega$ is varied. These
topological transitions can be quite complex with the Chern number
changing by $\pm 2,\pm 3$.
As discussed in Ref.~\onlinecite{Kundu14}, this occurs because
when linearly dispersing Dirac bands cross,
the Chern number exchanges between $\pm 1$, while quadratically dispersing band-crossings
cause the Chern number to exchange between $\pm 2$, and their combined
effect can lead to the topological transitions observed here and in Ref.~\onlinecite{Kundu14}.

Fig.~\ref{fig1a2} shows that the Hall conductance for the closed system after a quench is smaller
than that for the ideal case, this is not surprising as a quench creates a nonequilibrium population
of electrons which for a closed system of non-interacting electrons, has no means to relax.
The symmetry of the system dictates that the quasi-energies are located
symmetrically about zero.
An intriguing effect that can
occur is a reversal of the sign of the Hall conductance due to a laser-like situation where
the population in the ``upper'' quasi-band is higher. These populations are determined
by the overlap of the initial wavefunction and the Floquet modes, thus as $A_0at_h/\Omega$ is
varied, this overlap can be higher with one quasi-band or the other, leading to a reversal in the
sign of the Hall conductance that does not necessarily follow the sign of $C$. This phenomena was also
noticed in Ref.~\onlinecite{Rigol14}.

To highlight this effect, the excitation density $\rho_{kd}-\rho_{ku}$ that enters in the Hall conductance is plotted
in Fig.~\ref{fig3a4} for two different cases.
The upper panel of Fig.~\ref{fig3a4} is for the case where the initial wavefunction has the higher overlap
with the lower (or negative energy) Floquet band so that
the Hall conductance is the same sign as the ideal case, while the lower panel is for a case when
the initial wavefunction has a larger overlap with the
upper (positive energy) Floquet band so that the Hall conductance has the opposite sign to the ideal case.
Also a very general feature of the excitation density are spikes or enhanced excitations at the Dirac points.
We will show in the next section that this feature will persist even for the open system,
though the spikes will broaden as the temperature of the reservoir is increased.

Another feature one finds is that the Hall conductance after a quench shows jumps
that sometimes follow the topological transitions governed by jumps in $C$, but not always.
For example, in the upper and lower panel of Fig.~\ref{fig1a2}, one finds a topological transition
at $A_0a\sim 2.5$ where
the Chern number changes from $1 \rightarrow -2 \rightarrow 1$ very rapidly. The
Hall conductance after the quench on the other hand is sensitive to the
first transition from $1\rightarrow -2$, but not to the second from $-2\rightarrow 1$.
A similar effect is seen in the lower panel in Fig.~\ref{fig1a2} where $\sigma_{xy}^{\rm quench}$
does not follow the topological transition at $A_0a\sim 1$.

The quench results presented
here are relevant to the experimental set-up in Ref.~\onlinecite{Esslinger14} where
a Floquet topological system was realized in a closed cold-atomic gas, and where transport
measurements were performed by tilting the system and observing the magnitude of the
transverse drift in time of flight measurements. Another relevant situation is ultra-fast
pump probe measurements in solids using pulse lasers, when measurements are done faster than
phonon relaxation times.

\begin{figure}
\begin{center}
\begin{tabular}{lll}
 \includegraphics[height=8cm,width=8cm,keepaspectratio]{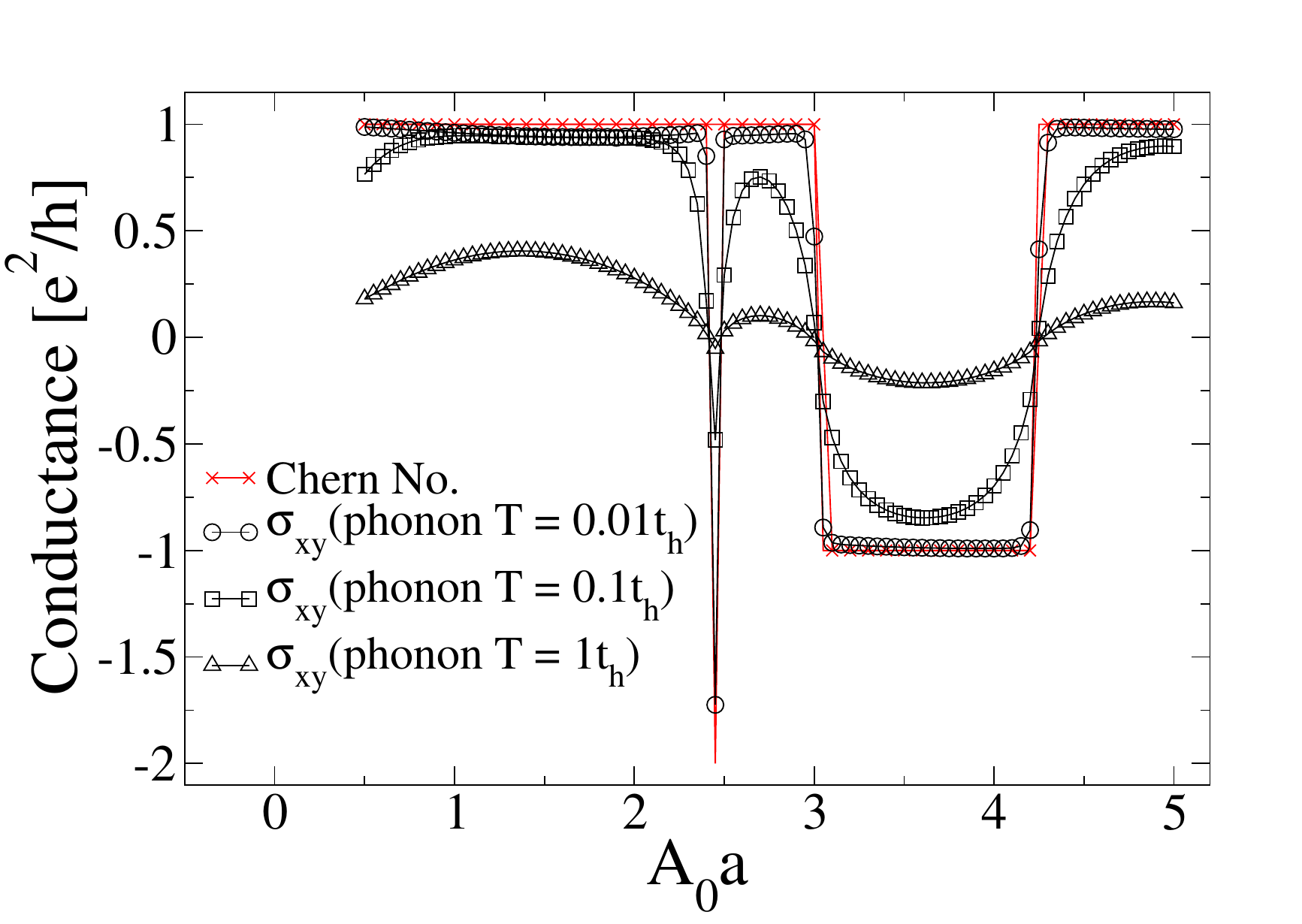}\\
(a)\\
\includegraphics[height=8cm,width=8cm,keepaspectratio]{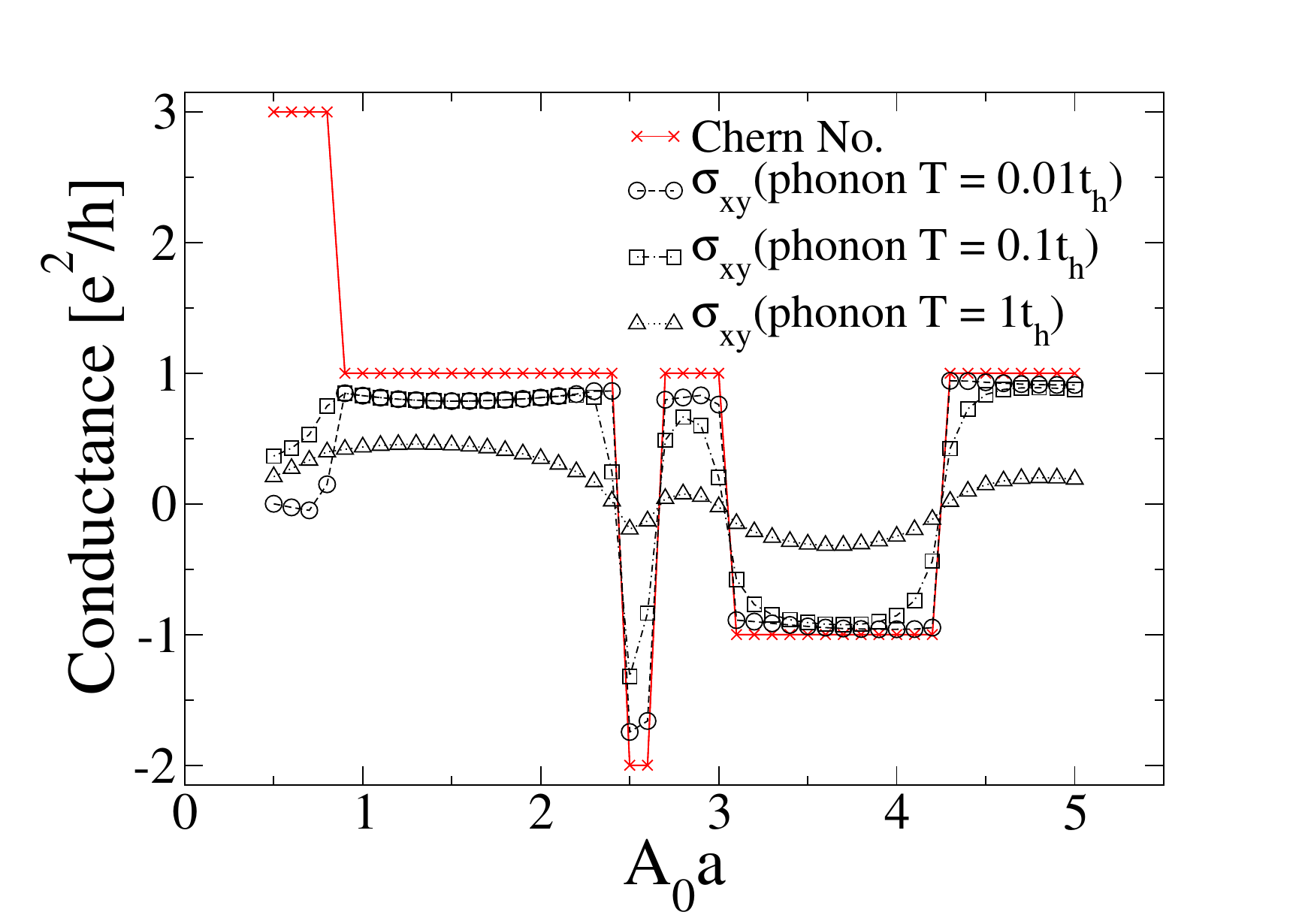}\\
(b)
\end{tabular}
\caption{(Color online) Hall conductance for the ideal case ($\sigma_{xy}= C e^2/h$)
and at steady state with a phonon reservoir, for different strengths of the circularly polarized laser and for
laser frequencies:
(a). $\Omega = 10t_h$ (b). $\Omega=5t_h$.
}
\label{fig5a6}
\end{center}
\end{figure}

\begin{figure}
\begin{center}
\begin{tabular}{lll}
 \includegraphics[height=8cm,width=8cm,keepaspectratio]{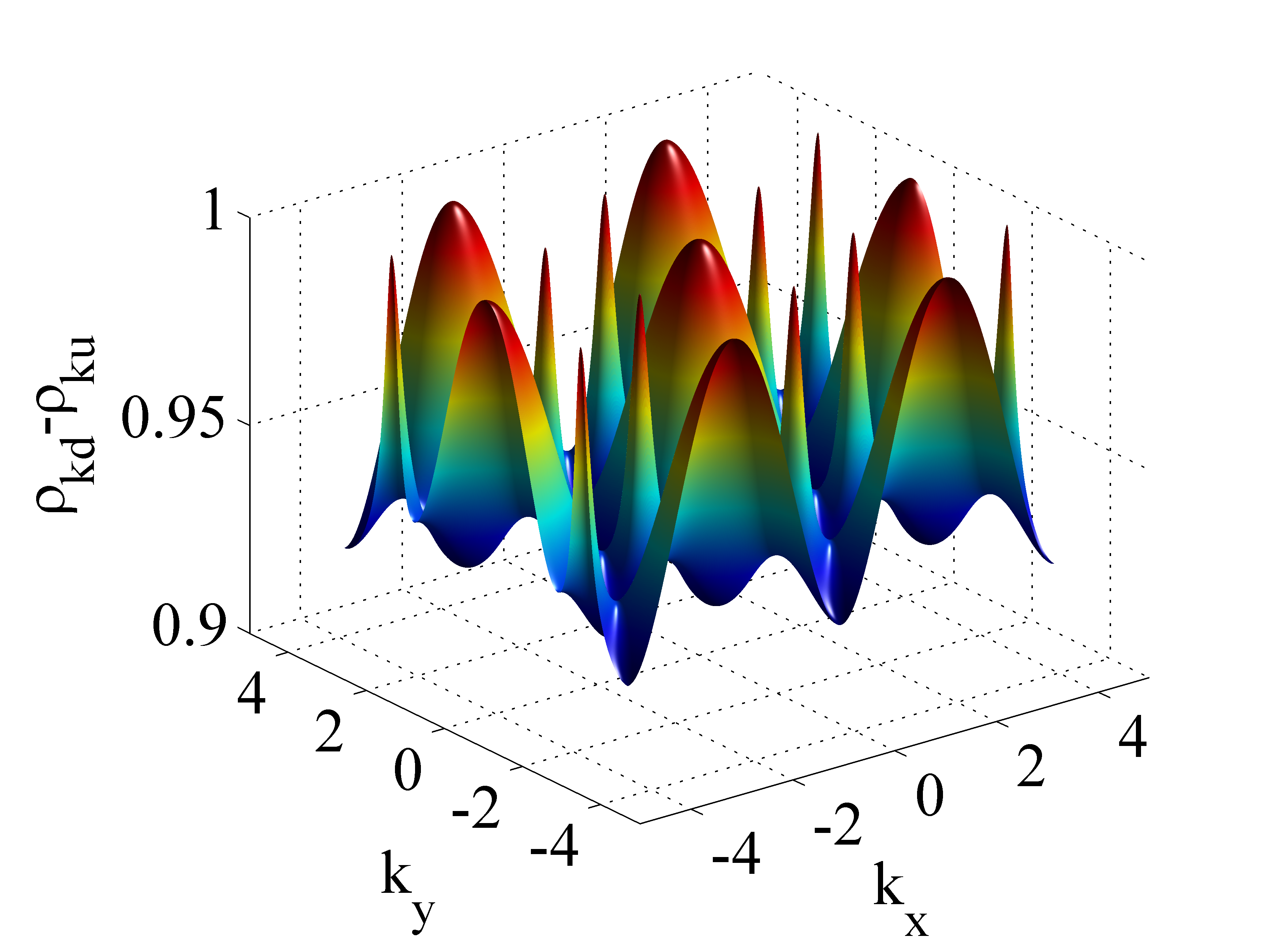}\\
(a)\\
\includegraphics[height=8cm,width=8cm,keepaspectratio]{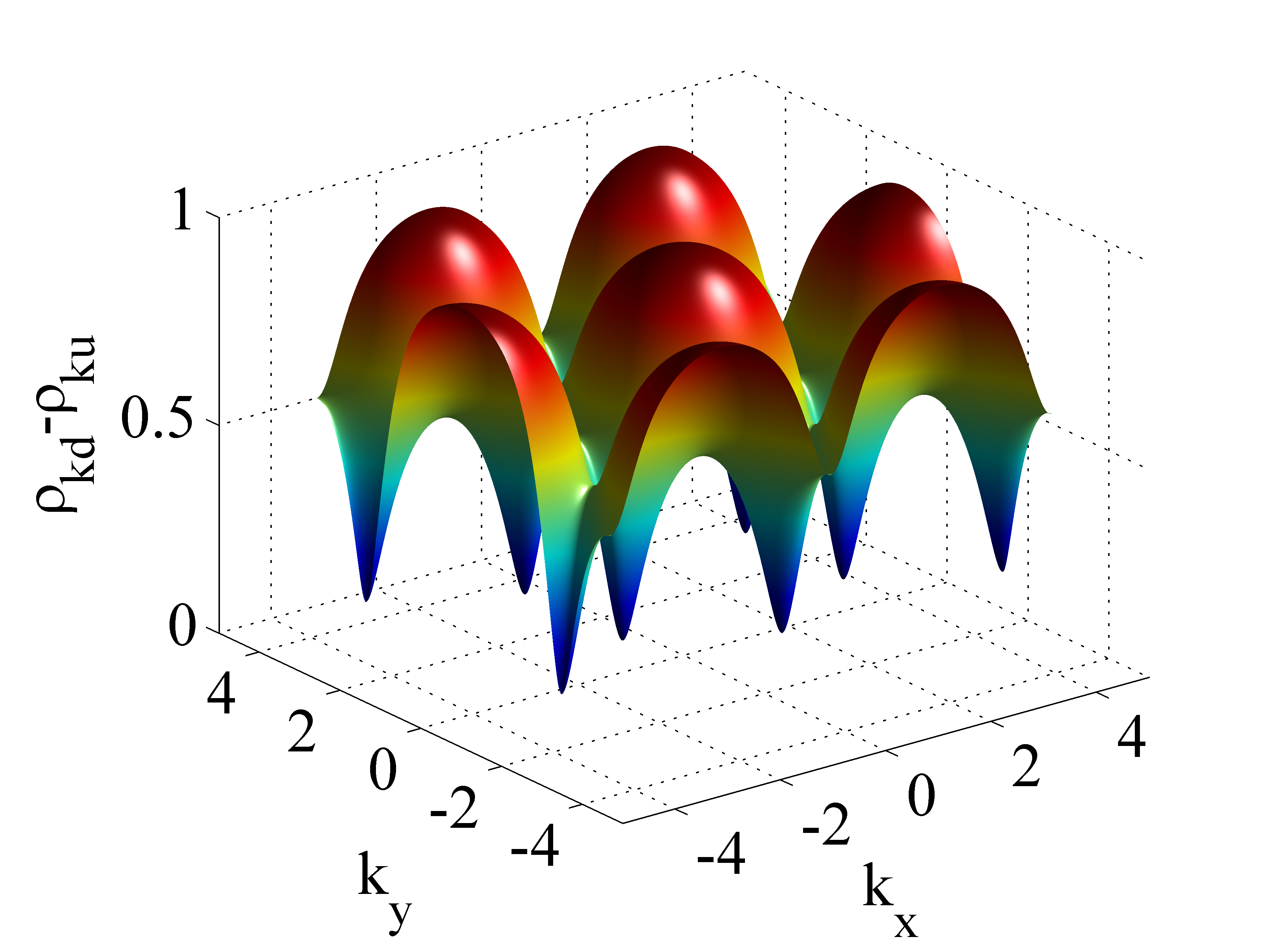}\\
(b)
\end{tabular}
\caption{(Color online) Excitation density $\rho_{kd}-\rho_{ku}$ at steady state with phonons.
The parameters are
$A_0a = 1.0, \Omega = 10t_h$, with the phonons at temperature (a) $T=0.01 t_h$, (b). $T=1.0t_h$.
}
\label{fig7a8}
\end{center}
\end{figure}

\section{Hall conductance for the open system}\label{phonons}
We now present results for the Hall conductance when the system is
coupled to an ideal reservoir of phonons that is always in
thermal equilibrium at a temperature $T$.
Inelastic scattering between electrons and phonons will cause the electron distribution
function to relax, affecting topological
properties such as the Hall conductance. We employ a rate or kinetic equation approach
within the Floquet formalism~\cite{Hanggi2005,Kohn01,Kohn09} to study how the electron distribution
evolves from an initial state generated by a quench switch-on protocol, and present analytic
results for the resulting steady-state.
A similar treatment was carried out for 2D Dirac fermions irradiated by a circularly
polarized laser and coupled to phonons,~\cite{Dehghani14} we generalize the approach of
Ref.~\onlinecite{Dehghani14} to graphene.

For completeness we first briefly outline the derivation of the kinetic equation.
Let $W(t)$ be the density matrix obeying
\begin{eqnarray}
\frac{dW(t)}{dt} = -i \left[H,W(t)\right]
\end{eqnarray}
It is convenient
to be in the interaction representation,
$W_{I}(t) = e^{i H_{\rm ph} t}U^{\dagger}_{\rm el}(t,0)W(t)U_{\rm el}(t,0)e^{- i H_{\rm ph} t}$,
where $U_{\rm el}(t,t')=\prod_kU_k(t,t')$ is the time-evolution operator
for the electrons under a periodic drive (see Eq.~(\ref{Udef})).
To ${\cal O}(H_{c}^2)$, the density matrix obeys the following equation of motion
\begin{eqnarray}
&&\frac{dW_I}{dt}=-i\left[H_{c,I}(t),W_I(t_0)\right]\nonumber\\
&&-\int_{t_0}^tdt'\left[H_{c,I}(t),\left[H_{c,I}(t'),W_{I}(t')\right]\right]
\end{eqnarray}
where
$H_{c,I}$ is in the interaction representation.
We assume that at the initial time $t_0$, the electrons and phonons are uncoupled so that
$W(t_0) = W^{\rm el}_0(t_0)\otimes W^{\rm ph}(t_0)$, and that
initially the electrons are in the post-quench state $|\Psi(t)\rangle$ described in Section~\ref{ideal}, while
the phonons are in thermal equilibrium at temperature $T$. This is justified because phonon
dynamics is much slower than electron dynamics, so that the quench state of section~\ref{ideal}
can be achieved within
femto-second time-scales,~\cite{Gedik13} while,
the phonons do not affect the system until pico-second
time-scales.

Thus,
\begin{eqnarray}
W^{\rm el}_0(t)= |\Psi(t)\rangle\langle \Psi(t)|=\prod_kW^{\rm el}_{k,0}
\end{eqnarray} where
\begin{eqnarray}
\!\!W^{\rm el}_{k,0}(t)=\!\!\! \sum_{\alpha,\beta=u,d}e^{-i(\epsilon_{k \alpha}-\epsilon_{k \beta})t}|\phi_{k\alpha}(t)\rangle
\langle\phi_{k\beta}(t)|
\rho_{k,\alpha\beta}^{\rm quench}
\end{eqnarray}
with
\begin{eqnarray}
\rho_{k,\alpha\beta}^{\rm quench}= \langle \phi_{k\alpha}(0)|\psi_{{\rm in},k} \rangle\langle\psi_{{\rm in},k}
| \phi_{k\beta}(0)\rangle
\end{eqnarray}

Defining the electron reduced
density matrix as the one obtained from tracing over the phonons,
$W^{\rm el} = {\rm Tr}_{\rm ph}W$, and noting that $H_c$ being linear in the phonon operators,
the trace vanishes, we need to solve,
\begin{eqnarray}
\frac{dW^{\rm el}_I}{dt}=-{\rm Tr}_{\rm ph}\int_{t_0}^tdt'\left[H_{c,I}(t),\left[H_{c,I}(t'),W_{I}(t')\right]\right]
\end{eqnarray}
We assume that the phonons are an ideal reservoir and stay in equilibrium. In that case $W_I(t) = W^{\rm el}_I(t)\otimes
e^{-H_{\rm ph}/T}/{\rm Tr}\left[ e^{-H_{\rm ph}/T}\right]$ (we set $k_B=1$).

The most general form of the reduced density matrix
for the electrons is
\begin{eqnarray}
W^{\rm el}_I(t)  =\prod_k \sum_{\alpha\beta}\rho_{k,\alpha \beta}(t) |\phi_{k\alpha}(t)\rangle\langle\phi_{k\beta}(t)|
\end{eqnarray}
where in the absence of phonons, $\rho_{k,\alpha \beta}=\rho_{k,\alpha\beta}^{\rm quench }$ and are time-independent
in the interaction representation.
The last remaining assumption is to identify the slow
and fast variables, which allows one to make the Markov approximation.~\cite{Hanggi2005}
We assume that $\rho_{k,\alpha\beta}$
are slowly varying as compared to the characteristic time scales of the reservoir. We also make
the so called modified rotating wave approximation~\cite{Kohn01} where it is assumed that the density
matrix $\rho_{k,\alpha\beta}$ varies slowly over one cycle of the laser. The last approximation is not necessary,
and was not made in Ref.~\onlinecite{Dehghani14}, where it was observed that indeed the density matrix varies slowly over one cycle
of the laser for sufficiently weak coupling to the reservoirs.

We only study the diagonal components of $\rho_{k,\alpha\alpha}$, which after the Markov approximation,
obey the rate equation
\begin{eqnarray}
\dot{\rho}_{k,\alpha\alpha}(t)
=-\sum_{\beta=u,d}L^k_{\alpha\alpha;\beta\beta}
\rho_{k,\beta\beta}(t)\label{rate}
\end{eqnarray}
$L^k_{\alpha\alpha,\beta\beta}$ are the in-scattering and out-scattering rates which due to
conservation of particle number obey $\sum_{\alpha=u,d}L^k_{\alpha\alpha,\beta\beta}=0$.

Thus to summarize, the main
approximations made in deriving Eq.~(\ref{rate}) are~\cite{Kohn09}, a). the phonon bath is
always in thermal equilibrium,
b). the system-bath coupling is weak as compared to  the laser frequency as well
as the bath relaxation rates,
c). the bath correlation times are short as compared to the time-scales over which
the reduced density matrix for the electrons varies, d). a
modified rotating wave approximation has been made where the scattering matrix
elements are replaced by their average over one cycle of the laser. This
is valid when the reduced density matrix varies slowly over one cycle of the laser, which
is typically the case when the system-bath coupling is weak in comparison to the laser
frequency.~\cite{Dehghani14}
The Floquet kinetic equation fully takes into account the time-periodic structure
of the Floquet states. The reduced density matrix components $\rho_{k,\alpha\alpha}$
are the occupation probabilities of these Floquet states, and it is these probabilities
that are assumed to be sufficiently slowly varying in time.

While the physical initial condition corresponds to a quench switch on protocol
for the laser where $\rho_{k,\alpha\alpha}(t=0)=\rho_{k,\alpha=u,d}^{\rm quench}$,
the steady-state solution is independent of this initial state and corresponds to $\rho_{k,\alpha\alpha}(t=\infty)=\rho_{k\alpha}$, where
\begin{eqnarray}
\rho_{ku}= \frac{|L^k_{uu,dd}|}{|L^k_{uu,dd}|+|L^k_{uu,uu}|}; \rho_{kd}=1-\rho_{ku}\label{dss}
\end{eqnarray}
Expanding $\langle\phi_{k\alpha}(t)|c_{k\sigma}^{\dagger}c_{k\sigma'}|\phi_{k\beta}(t)\rangle$ in a Fourier series such that
$\langle\phi_{k\alpha}(t)|c_{k\uparrow}^{\dagger}c_{k\downarrow}|\phi_{k\beta}(t)\rangle=\sum_ne^{in\Omega t}C^n_{1k,\alpha\beta}$, and $C^n_{2k,\alpha\beta}
= \left(C^{-n}_{1k,\beta\alpha}\right)^*$,
we find the following in-scattering and out-scattering rates for a uniform phonon density of states $D_{\rm ph}$,
\begin{widetext}
\begin{eqnarray}
&&L^k_{uu,uu} = D_{\rm ph}\biggl[\lambda_x^2\sum_n\left(C^{n}_{1k,ud}C^{-n}_{1k,du}+ C^{n}_{2k,ud}C^{-n}_{2k,du}+C^{n}_{1k,ud}C^{-n}_{2k,du}
+C^{n}_{2k,ud}C^{-n}_{1k,du}\right) \nonumber\\
&&+ \lambda_y^2\sum_n\left(-C^{n}_{1k,ud}C^{-n}_{1k,du}- C^{n}_{2k,ud}C^{-n}_{2k,du}+C^{n}_{1k,ud}C^{-n}_{2k,du}
+C^{n}_{2k,ud}C^{-n}_{1k,du} \right)\biggr]\nonumber\\
&&\times \biggl[\theta(-\epsilon_{kd}+\epsilon_{ku}+n\Omega)(1+N(-\epsilon_{kd}+\epsilon_{ku}+n\Omega))
+\theta(\epsilon_{kd}-\epsilon_{ku}-n\Omega)N(\epsilon_{kd}-\epsilon_{ku}-n\Omega)\biggr]\label{uuuu}\\
&&-L^k_{uu,dd} = D_{\rm ph}\biggl[\lambda_x^2\sum_n\left(C^{n}_{1k,ud}C^{-n}_{1k,du}+ C^{n}_{2k,ud}C^{-n}_{2k,du}+C^{n}_{1k,ud}C^{-n}_{2k,du}
+C^{n}_{2k,ud}C^{-n}_{1k,du}\right) \nonumber\\
&&+ \lambda_y^2\sum_n\left(-C^{n}_{1k,ud}C^{-n}_{1k,du}- C^{n}_{2k,ud}C^{-n}_{2k,du}+C^{n}_{1k,ud}C^{-n}_{2k,du}
+C^{n}_{2k,ud}C^{-n}_{1k,du} \right)\biggr]\nonumber\\
&&\times \biggl[\theta(\epsilon_{kd}-\epsilon_{ku}-n\Omega)(1+N(\epsilon_{kd}-\epsilon_{ku}-n\Omega))
+\theta(-\epsilon_{kd}+\epsilon_{ku}+n\Omega)N(-\epsilon_{kd}+\epsilon_{ku}+n\Omega)\biggr]\label{uudd}
\end{eqnarray}
\end{widetext}
Above $N(x) = 1/(e^{x/T}-1)$ is the Bose function.
In presenting our results we also consider an isotropic electron-phonon coupling $\lambda_x=\lambda_y$ so that the steady-state
electron distribution function becomes independent of the electron-phonon coupling.
\begin{figure}
\begin{center}
\begin{tabular}{lll}
 \includegraphics[height=8cm,width=8cm,keepaspectratio]{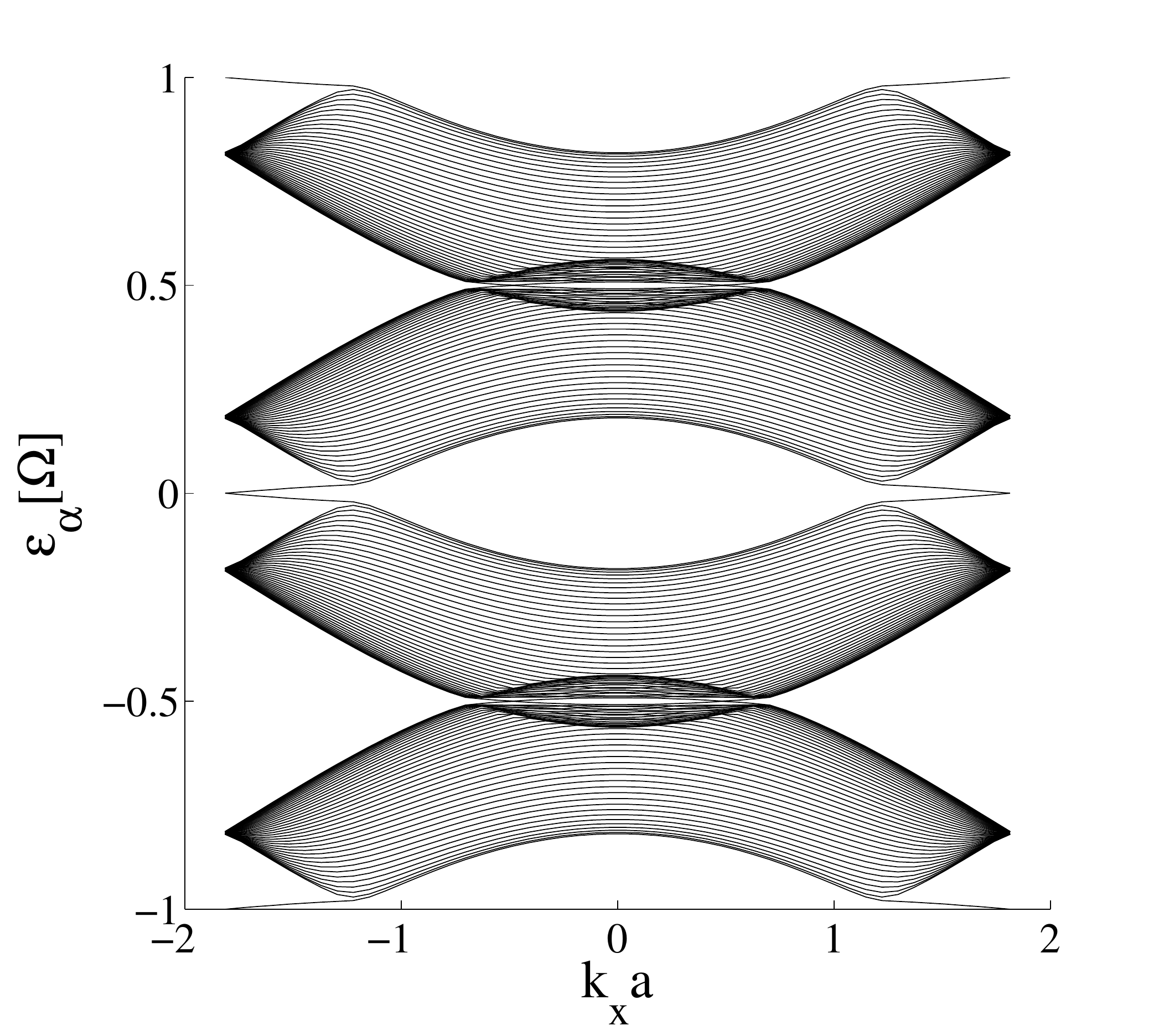}\\
(a)\\
\includegraphics[height=8cm,width=8cm,keepaspectratio]{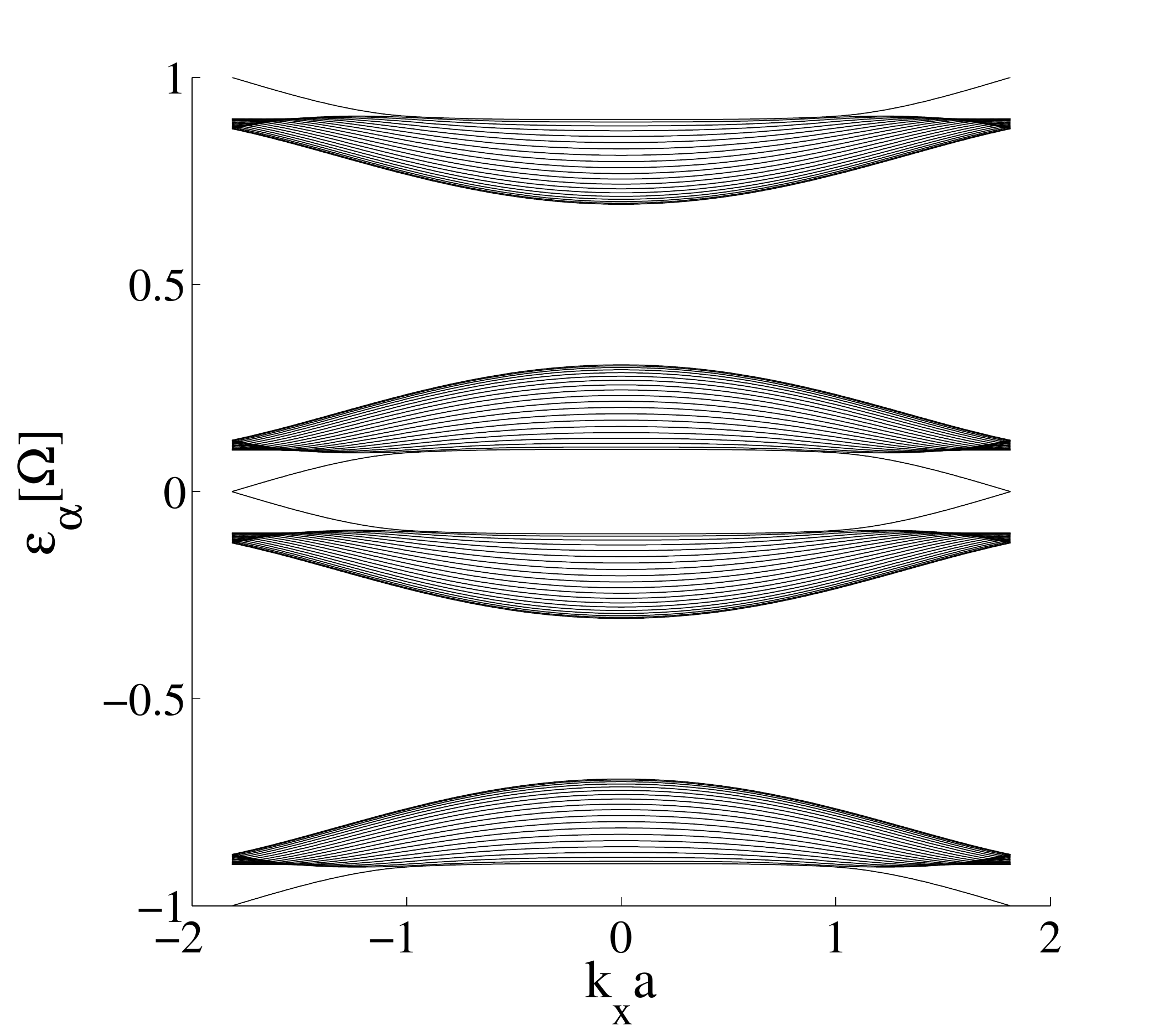}\\
(b)
\end{tabular}
\caption{(Color online) Floquet spectrum over two Floquet BZs
for laser frequency $\Omega=5.0 t_h$ and for laser amplitudes and Chern numbers,
(a). $A_0a = 0.5, C=3$ (b). $A_0a = 1.5,C=1$. Additional edge states at the Floquet zone boundaries appear for case (a).
}
\label{fig11a12}
\end{center}
\end{figure}

Eqs.~(\ref{uuuu}) and ~(\ref{uudd}) imply that the population of the two quasi-bands $\rho_{kd,u}$ are determined by
a sum over phonon induced inelastic scattering between many quasi-energy levels (denoted by the sum over
$n$). These complicated scattering processes imply a nonequilibrium (non-Gibbsian) steady-state
for the electrons even when the phonons are in thermal equilibrium, unless the frequency of the laser is so high that
only a single term in the sum over $n$ survives~\cite{Dehghani14,Shirai14}. In such a high-frequency limit,
as we shall show, the Hall conductance approaches a thermal result, and in particular will approach $C e^2/h$
as the reservoir temperature is lowered. For lower laser frequencies on the other hand, significant deviations
from $C e^2/h$ will be found even when the phonons are at a very low temperature.

Fig.~\ref{fig5a6} shows the steady-state Hall conductance for three different reservoir temperatures ($T=0.01t_h, 0.1 t_h,1 t_h$), and for
the same laser parameters as the ones for which the quench results were discussed. These results are plotted with those for the
``ideal'' case. Fig.~\ref{fig5a6} (a) is for a fairly high frequency ($\Omega=10t_h$) and shows
that the steady-state Hall conductance approaches the ideal limit
of $C e^2/h$ as the temperature of the reservoir is lowered, with the topological transitions
characterized by a thermal broadening. The excitation density for the same laser frequency is shown in Fig.~\ref{fig7a8}, and
is characterized by sharp spikes at the Dirac points at
low temperatures which then show thermal broadening as the
temperature of the reservoir is raised.

Fig.~\ref{fig5a6} (b) is for a lower laser frequency of $\Omega= 5 t_h$. In this case, while for large laser amplitudes ($A_0a>1$),
the results are similar to panel (a), with the Hall conductance approaching $Ce^2/h$ as the temperature of the bath is lowered,
marked deviations are seen for smaller laser amplitudes ($A_0a<1$). For this case the
Hall conductance, even with low temperature phonons, saturates at a value very different from $Ce^2/h$, infact almost
approaching zero.

Even though we have a large sample in mind, where the role played by the
edges do not explicitly enter the calculation, it is still instructive to
study the quasi-energy spectrum in a finite geometry (Fig.~\ref{fig11a12}) to understand the difference between
the case of $A_0a>1$ and $A_0a<1$, but at the same laser frequency $\Omega = 5 t_h$.
One observes that $A_0a>1$ is also the case where the laser frequency is large as
compared to the electron band-width (which is strongly influenced by $A_0a$), and all the edge-states reside at the center of the
Floquet BZ ($\epsilon=0$), with the number of chiral edge modes equaling the Chern number
$C$. In contrast for laser frequencies comparable to or smaller
than the band-width, ($A_0a<1$), additional edge modes appear in the Floquet zone boundaries ($\epsilon =\pm\Omega/2$),
and the number of chiral edge modes no longer equal the Chern number $C$, which is no longer a good or
sufficient topological index.
A modified topological invariant has been introduced that correctly counts
the number of edges modes at the center and edges of the zone-boundary~\cite{Rudner13,Carpentier14}, however
we find that the distribution function at low frequencies is so far out of equilibrium, that the
Hall conductance is unrelated to this new topological invariant, and almost approaches zero. Thus highly nonequilibrium
steady-states for small laser frequencies prevent one to achieve Hall conductances of ${\cal O}(e^2/h)$.

\begin{figure}
\begin{center}
\begin{tabular}{lll}
 \includegraphics[height=8cm,width=8cm,keepaspectratio]{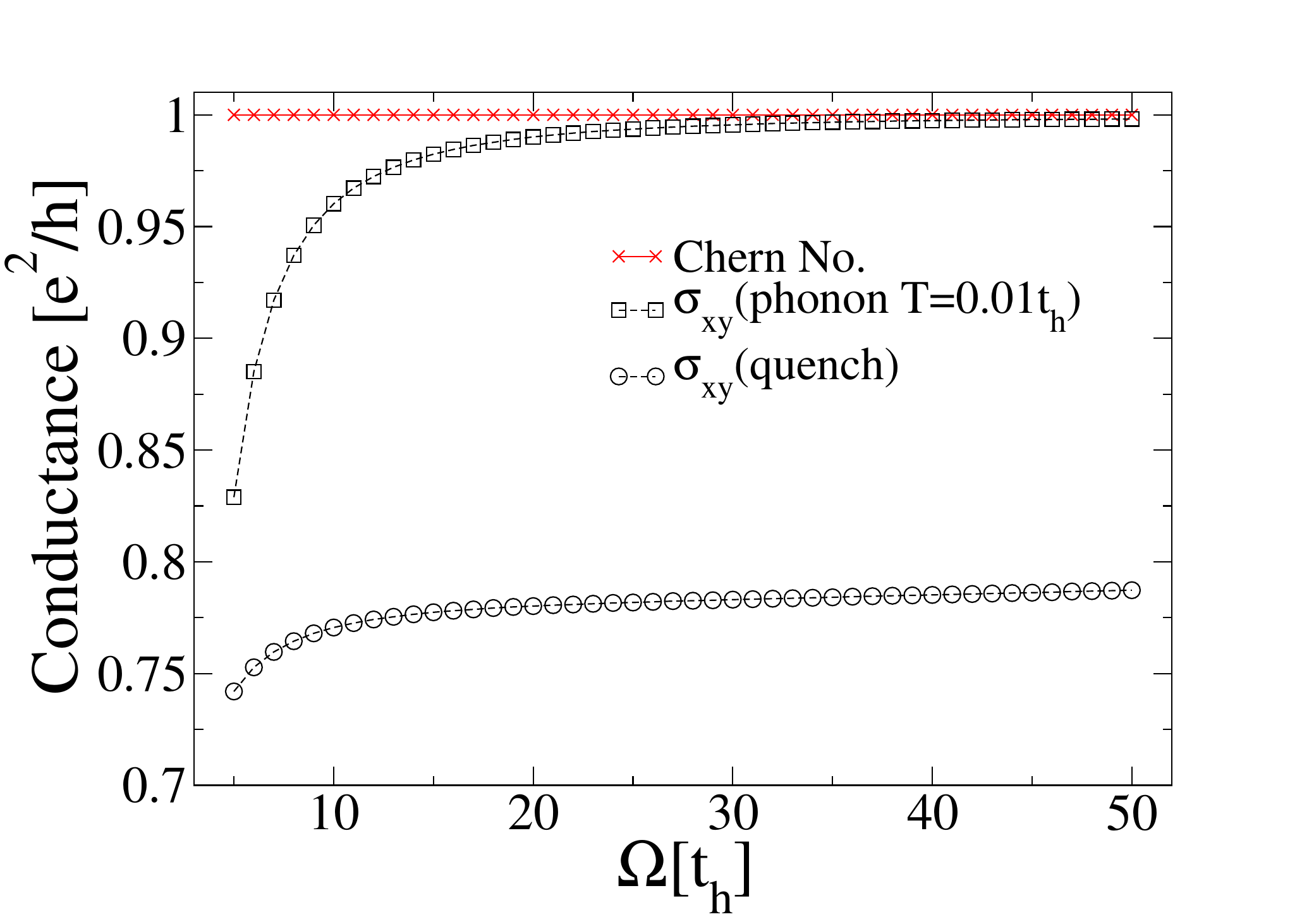}\\
(a)\\
\includegraphics[height=8cm,width=8cm,keepaspectratio]{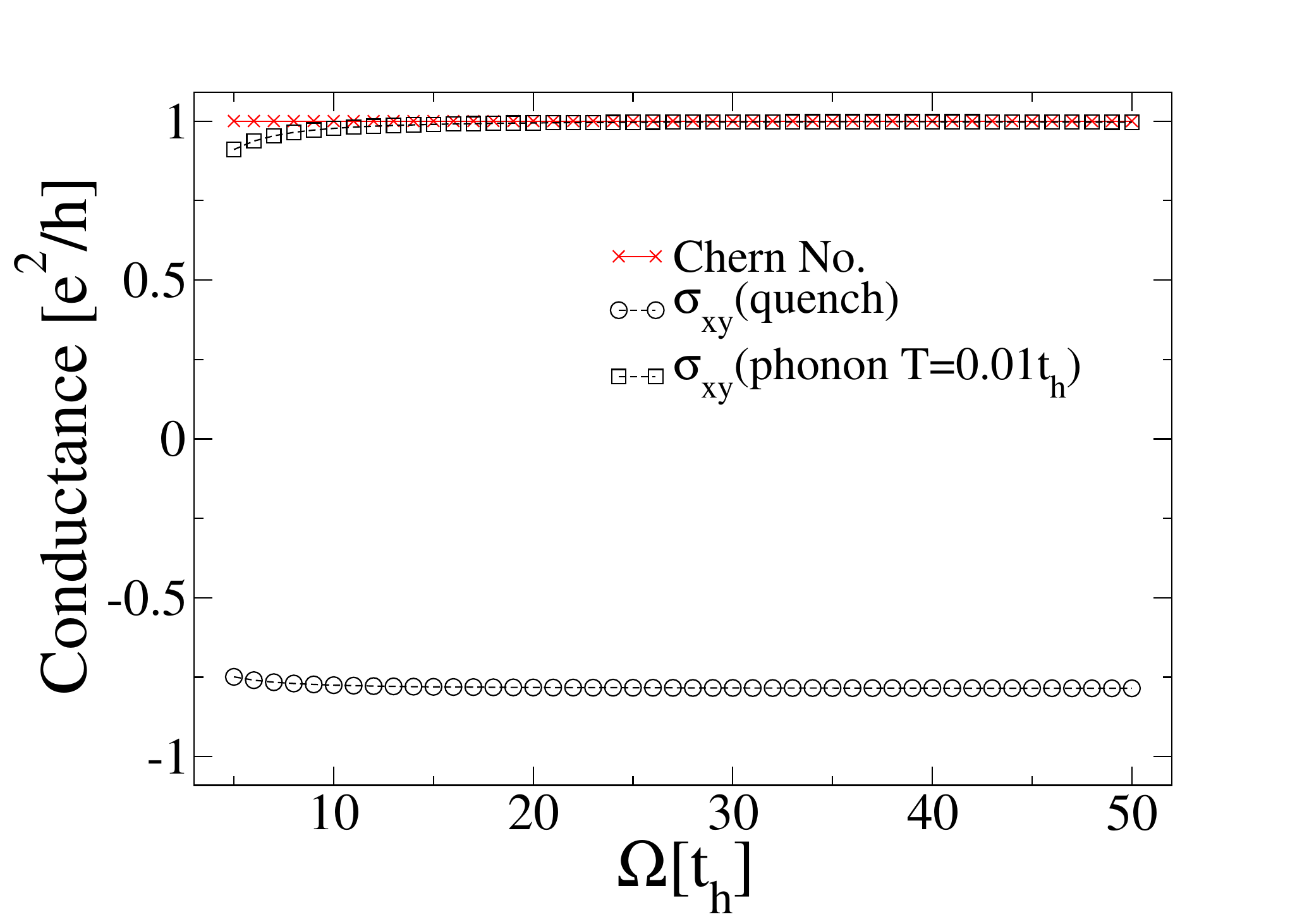}\\
(b)
\end{tabular}
\caption{(Color online) Hall conductance for the closed system after a quench, for the open system at steady-state
with phonons at $T=0.01 t_h$, and for the ideal case ($C e^2/h$), plotted for different laser frequencies and for the
laser amplitudes:
(a). $A_0a = 1.0$ (b). $A_0a = 5.0$.
}
\label{fig9a10}
\end{center}
\end{figure}

Fig.~\ref{fig9a10} shows how the Hall conductance depends upon the frequency of the laser for the closed as
well as for the open system, where for the latter the reservoir temperature is fairly low ($T=0.01t_h$).
As the laser frequency is increased,
the Hall conductance for the open system approaches the ideal quantum limit, and the results become more
and more like an equilibrium system where the Floquet bands are occupied by the Gibbs distribution.~\cite{Dehghani14,Shirai14}
The closed system of course corresponds to a nonequilibrium situation as there is no mechanism for thermalization, with
the steady-state depending on the overlap between the initial state and the Floquet state, resulting in a Hall conductance
that can have the opposite sign to that of the open system.

\section{Conclusions}\label{conclu}

We have studied the dc Hall conductance derived from the Kubo formula, for
graphene irradiated by a circularly polarized laser. Results are presented
for two situations, one is for a closed system for a quench switch-on protocol for
the laser, while the second is for an open system coupled to
an ideal phonon reservoir.
For the closed system, the electron distribution function retains memory of the initial conditions which
can lead to Hall conductances (Fig.~\ref{fig1a2}) that are not only smaller in magnitude than the ideal limit of $Ce^2/h$, but also
sometimes do not follow the topological transitions in $C$ as the laser parameters are varied, and can
be of the opposite sign to the ideal result. The latter occurs when the
initial state has a larger overlap (Fig.~\ref{fig3a4}) with the ``upper'' Floquet band
which has a Berry curvature of the opposite sign to that of the ``lower'' Floquet band. The results for
the closed system are most relevant for experiments in cold-atomic gases such as the one of Ref.~\onlinecite{Esslinger14}.

For the open system, as long as the
laser frequencies are larger than the electron bandwidth (for small laser amplitudes $A_0a<1$,
this condition is $\Omega > 6t_h$), the main effect of the reservoir
is to cause an effective cooling that allows the Hall conductance to eventually approach $Ce^2/h$ as the reservoir
temperature is lowered (Fig.~\ref{fig5a6}, upper panel and Fig.~\ref{fig9a10}), with the Hall conductance following
the topological transitions with a characteristically thermal broadening.

For the open system, surprises occur for laser frequencies lower or comparable to the band-width (Fig.~\ref{fig5a6} lower panel).
In this case, strong deviations of the Hall conductance from $C e^2/h$ occur, with the Hall conductance almost
approaching zero. This may be related to the Hall transport measured in graphene irradiated by
THz laser,~\cite{Karch10,Karch11} where the observed Hall effect was very small compared to the quantum limit,
and was accounted for by a semi-classical Boltzman analysis.

Interestingly enough these strong deviations from the quantum limit
are also accompanied by the appearance of edge-states in the BZ edges so that $C$ is no longer a good
topological index. However the result we obtain cannot be accounted for by any modified topological index that
takes into account these new edge modes. This is because,
the electron distribution function for low laser frequencies
is highly out of equilibrium even when the reservoir is ideal, with the resultant steady-state determined from
solving a rate equation that accounts for
laser induced photoexcitation of carriers and phonon induced inelastic scattering between many different
quasi-energy levels.

These results also suggest that due to the inherent nonequilibrium nature of the problem, especially
for low laser frequencies,
the Hall conductance will depend upon the dominant inelastic scattering mechanism, and hence the Hall conductance
in large samples where electron-phonon scattering is dominant will differ from the Hall conductance in smaller
samples~\cite{Torres14} where for the latter the
relaxation mechanism is determined by the location of the Fermi-levels of the leads.~\cite{Fertig11,Kundu14,Torres14}
It is of course interesting to also consider samples of intermediate size where both the leads as well
as the phonons play a role in the inelastic scattering.~\cite{Lindner15}

For the experimental
feasibility of observing a large Hall response of ${\cal O}(e^2/h)$, one therefore needs laser frequencies larger than the electron band-width
as this suppresses photoexcited carriers, and eliminates edge-states
at the Floquet BZ boundaries making $C$ the relevant topological index. However,
one needs to keep in mind that
the maximum voltage drop across a lattice site due to the applied laser
($\sim A_0 a \Omega$) cannot be too large in order to avoid dielectric breakdown across orbital sub-bands, at the same time
the laser amplitude $A_0a$ should be large enough so that
the dynamical gap $\Delta$ at the Dirac points is larger than the temperature of the reservoir. With current day experiments,
one may realize these conditions in artificial graphene lattices such as in cold-atomic gases~\cite{Esslinger14}
and photonic waveguides.~\cite{Segev13}

{\sl Acknowledgments:}
This work was supported by US Department of Energy, Office of Science, Basic Energy Sciences,
under Award No. DE-SC0010821 (HD and AM).

%

\end{document}